%% file: entire_manuscript.tex
\documentclass[
 reprint,
 doublecolumn,
 amsmath,amssymb,
 aps,
 prx
]{revtex4-2}

\usepackage{graphicx} 
\usepackage{amsmath}
\usepackage{booktabs} 
\usepackage{physics}
\newcommand{\dressedstate}[1]{\ket{\overline{#1}}}

\begin{document}

\title{Unlocking a fast adiabatic CZ gate and exact residual $ZZ$ cancellation between fixed-frequency transmons using a floating tunable coupler 
}
\input{authors_affiliations.tex}
\date{\today}

\makeatletter
\let\saved@maketitle\maketitle
\makeatother

\begin{abstract}
Tunable couplers in superconducting qubit architectures enable strong qubit-qubit interactions for two-qubit gates while suppressing unwanted coupling during single-qubit operations. However, achieving low error rates for fast two-qubit gates remains challenging since suppressing leakage and non-adiabatic errors typically requires specialized qubit, coupler, or pulse designs, often at the expense of an idling $ZZ=0$ condition.  In this work, we demonstrate that a symmetric floating tunable coupler provides a natural platform for fast, high-fidelity adiabatic controlled-Z (CZ) gates. Its favorable energy-level structure eliminates the conventional trade-off between rapid conditional-phase accumulation and adiabatic evolution while preserving exact cancellation of residual $ZZ$ interaction at idling. This architecture exhibits intrinsic robustness to non-adiabatic transitions, even under simple flux modulation waveforms. To push performance at short gate durations, where maintaining adiabaticity becomes more challenging despite the favorable level structure, we introduce pulse-shaping techniques based on the instantaneous adiabatic factor that further suppress non-adiabatic errors.
We experimentally realize a 24~ns adiabatic CZ gate with fidelity exceeding 99.9\% and stable operation over several hours with a time-averaged fidelity of 99.878\%.
\end{abstract}

\maketitle

\input{main.tex}

\clearpage
\appendix

\makeatletter
\let\maketitle\saved@maketitle
\makeatother

\title{Supplementary material for: Unlocking a fast adiabatic CZ gate and exact residual $ZZ$ cancellation between fixed-frequency transmons using a floating tunable coupler }
\input{authors_affiliations.tex}
\date{\today}

\maketitle

\input{supplementary.tex}

\bibliography{references.bib}

\end{document}

%% file: authors_affiliations.tex
\author{Angela Q. Chen}
\author{Sarah Strong}
\author{Xian Wu}
\author{Stefano Poletto}\thanks{email: stefano@rigetti.com}
\affiliation{
Rigetti Computing, 775 Heinz Avenue, Berkeley, CA 94710
}

%% file: main.tex
In recent years, superconducting quantum processors based on tunable couplers have emerged as a leading architecture for scalable quantum computing \cite{FeiYan2018,Sete2021,Arute2019,LiX2020,Sung2021,Marxer2023,Collodo2020,Xu2020,Stehlik2021,Mundada2019}.
By dynamically modulating the interaction between qubits, tunable couplers enable strong coupling during gate operation while maintaining minimal residual interaction at idle, without requiring a large qubit-qubit detuning.
This combination of isolation and controllability makes couplers particularly attractive for high-performing multi-qubit platforms.

Two-qubit gates in tunable-coupler architectures are typically implemented using either resonant (diabatic) interaction or adiabatic phase accumulation.
In resonant schemes, energy levels in the one- or two-excitation manifolds are intentionally aligned to induce population exchange or $\pi$-phase evolution via full swap with a non-computational state \cite{Majer2007,Yamamoto2010}.
While implementing these resonant gates on tunable coupler systems can be fast~\cite{Sung2021,Marxer2025,Arute2019,LiX2020}, they rely on full excitation transfer between computational and non-computational states, making them more susceptible to leakage and non-adiabatic errors.
The pulse shape of resonant gates must balance gate speed against unwanted spectral weight at avoided crossings of the system energy levels \cite{Sung2021,Marxer2025}, which is more challenging as the gates become faster.
To suppress this leakage, interferometry techniques were proposed for resonant gates enacted on fixed-coupling architectures \cite{Negrneac2021}. However, this method is most effective in systems with a single dominant leakage channel and does not naturally extend to the more complicated energy structure of tunable-coupler architectures, which have multiple leakage channels with distinct dynamical phases.

An alternative controlled-Z (CZ) gate scheme exploits the level hybridization of the $\ket{11}$ state with non-computational states. In this gate scheme, a controlled phase is accumulated as the $|11\rangle$ state adiabatically evolves \cite{Dicarlo2009,Barends2016}.
Previous demonstrations of adiabatic gates in tunable coupler architectures were implemented with grounded transmons \cite{Collodo2020, Xu2020,Glaser2025,VallesSanclemente2026}, which consist of one electrically floating electrode pad connected with a Josephson junction to the ground plane. However, because of the eigenstate ordering that arises in grounded transmon systems, achieving large phase accumulation while remaining in the instantaneous $\ket{11}$ eigenstate required operation in the non-straddling regime, where the qubit-qubit detuning exceeds the anharmonicity of the qubits. 
In this all-transmon architecture, unwanted residual $ZZ$ interaction at idling can be made small but not fully canceled in the non-straddling regime. 

More generally, previous studies of tunable-coupler systems have highlighted a fundamental trade-off between three key factors: large dynamical phase accumulation for fast gates, an energy-level ordering favorable for robust adiabatic evolution, and exact cancellation of the residual $ZZ$ interaction at idling~\cite{FeiYan2021}. In grounded tunable coupler architectures, fast gates are typically realized in the non-straddling regime because its energy-level configuration produces fewer anticrossings along the $|11\rangle$ trajectory \cite{Collodo2020,Xu2020,Glaser2025}. However, in the non-straddling regime, the residual $ZZ$ is negative near zero effective exchange coupling and decreases further with increasing effective exchange-coupling strength, preventing a $ZZ = 0$ crossing~\cite{FeiYan2021}. As a result, the operating regime that favors adiabaticity and fast gates in grounded coupler architectures does not coincide with a $ZZ = 0$ condition. More complex coupler implementations can alleviate the constraints imposed by the energy-level ordering of grounded couplers and asymmetric floating couplers (both qubits coupled predominantly to the same superconducting pad of the coupler), but at the cost of increased fabrication complexity and control overhead~\cite{Goto2022, GotoNakamura2024, An2025}. Furthermore, grounded tunable coupler architectures, including an early implementation with a grounded coupler and floating qubits~\cite{Stehlik2021}, rely on a direct qubit-qubit coupling capacitance to reach zero-coupling. This approach constrains qubit-qubit spacing, which limits scalability \cite{Marxer2023,Field2024}.

In this work, we show that the symmetric floating tunable coupler, in which the coupler pads are arranged symmetrically with respect to the qubits \cite{Sete2021}, provides a natural platform for fast, high-fidelity CZ gates while preserving a $ZZ=0$ condition. The gate is implemented in a simple and scalable architecture, requiring no additional fabrication complexity, and is activated by a single flux-pulse applied to the tunable coupler, thereby eliminating the need for tunable qubits and associated qubit flux noise. A schematic of the qubit-symmetric coupler-qubit system is shown in the inset of Fig.~\ref{fig:sim_sym}(a). From an analysis of the coupler’s energy-level structure, we identify how its ordering enables simultaneous access to the $ZZ=0$ condition and adiabatic gate trajectories with large phase accumulation. Guided by this analysis, we use quantified adiabatic factors to experimentally realize a 24\,ns adiabatic CZ gate with a fidelity as high as 99.927$\pm$0.015\%. These results demonstrate that fast and high-performing adiabatic gates in an interaction-free idling system can be achieved without added design complexity or constraints on scalability.

\section{Device simulations}
\begin{figure}[htbp]
    \centering
    \includegraphics[width=\columnwidth]{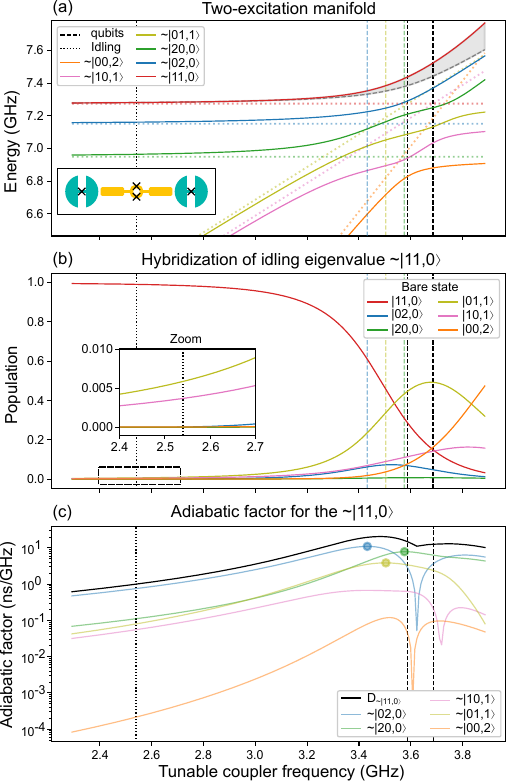}
    \caption{Simulated properties versus tunable coupler frequency for the symmetric floating coupler configuration. Vertical black dashed and dotted lines indicate, respectively, qubit frequencies and coupler idling frequency for exact cancellation of residual $ZZ$. Colored lines share the same color-scheme in the plots for easy comparison.
    (a) Two-excitation manifold energy levels labeled with their closest bare eigenvector at idling. Solid lines indicate the adiabatic trajectory of the dressed state; dotted lines represent the uncoupled energy states. The dashed gray line is the reference $E_{10,0}+E_{01,0}-E_{00,0}$ and the gray area denotes the dynamical phase accumulation rate $\zeta$. A device schematic is shown in the inset, with fixed-frequency qubits (teal) connected by the symmetric coupler (gold).
    (b) Hybridization of the state $\dressedstate{11,0}$ with bare two-excitation states. Inset is a zoom-in of the hybridization near idling.
    (c) Adiabatic factor for $\dressedstate{11,0}$ over the five different two-excitation states. The black solid line is the sum over all five contributions. Colored dots represent the position of the maximum value of the main adiabatic factors. These frequency locations are represented in (a) and (b) with colored vertical dashed lines.
    }
    \label{fig:sim_sym}
\end{figure}

In the transmon approximation and with $\hbar=1$, the coupler system is described by the Hamiltonian
\begin{align*}
    H = &\sum_{k=1,2,c}\left(\omega_k a_k^\dagger a_k + \frac{\eta_k}{2}a_k^\dagger a_k^\dagger a_k a_k \right) \\
      & + \sum_{j=1,2} g_{jc}\left(a_j^\dagger + a_j\right) \left(a_c^\dagger + a_c\right) \\
      & + g_{12}\left(a_1^\dagger + a_1\right) \left(a_2^\dagger + a_2\right),
\end{align*}
where $a_k^\dagger$ and $a_k$ are creation and annihilation operators for qubit $k=1,2$ and coupler $c$, $\omega_k/2\pi$ and $\eta_k/2\pi$ denote their bare frequencies and anharmonicities, $g_{jc}$ the qubit-coupler coupling, and $g_{12}$ the direct qubit-qubit coupling.
In capacitively coupled circuits, the interaction strengths scale with the mode frequencies as $g_{ij}=\rho_{ij}\sqrt{\omega_i\omega_j}$, where $\rho_{ij}$ is a frequency-independent parameter determined by the circuit capacitance (supplementary material \ref{sec:g2rho}).
In the symmetric floating configuration, the direct coupling $g_{12}$ and the interaction mediated by the coupler contribute with opposite signs, leading to an effective $g_{\mathrm{eff}}=0$ condition when the coupler frequency is below the qubit frequencies as demonstrated previously in Ref.~\cite{Sete2021}.

In Fig.~\ref{fig:sim_sym}, the energy profile, hybridization, and adiabatic factors of the system are simulated with QuTiP~\cite{qutip} using three energy levels for each qubit and the tunable coupler. We simulate the real device using the experimental parameters measured and extracted in Section~\ref{sec:device_characterization}. Energy levels and states are labeled as $E_{q_1q_2,TC}$ and $|q_1q_2, TC\rangle$, respectively, with $\omega_1<\omega_2$.
When the tunable coupler is in its ground state, the qubit subspace is denoted using the compact notation $\ket{q_1q_2}$ for clarity.

In the straddling regime of operation, the two-excitation computational state $|11\rangle$ lies at a higher energy than the second excited state of each qubit.
In this regime, the idling point corresponding to exact cancellation of residual $ZZ$ interaction occurs for tunable coupler frequencies below the qubit frequencies (illustrated by the black, dotted vertical line in Fig.~\ref{fig:sim_sym}(a) and demonstrated experimentally in Fig.~\ref{fig:freqs_and_zz}(b)). 
The adiabatic gate is implemented by flux-pulsing the tunable coupler toward the qubit frequencies while ensuring that the system remains on its adiabatic trajectory.
Along this trajectory, states are labeled by the bare state to which they are closest at the idling point and are denoted by $\dressedstate{q_1q_2,TC}$ to distinguish them from the pure bare states.

The dynamical phase rate $\zeta$---accumulated by the adiabatic trajectory of the $|11\rangle$ state relative to the reference energy $E_{10} + E_{01} - E_{00}$---enables the implementation of a controlled-phase gate from engineering the frequency excursion of the tunable coupler.
This $\zeta$-rate, illustrated by the gray region in Fig.~\ref{fig:sim_sym}(a), keeps growing in an unbounded manner with coupler frequency (simulation of $\zeta$ is included in Section~\ref{sec:supplement_symVasym} of the supplement). In addition to the unbounded increase of $\zeta$, we also highlight that the adiabatic trajectory of the $|11\rangle$ state does not pass through any anticrossings nor approach small energy gaps, which makes this operating regime particularly favorable for adiabatic evolution.

The hybridization of the qubit states is dynamically modulated by the tunable coupler frequency (Fig.~\ref{fig:sim_sym}(b)).
At the idling point, the computational state $\dressedstate{11,0}$ is partially hybridized with tunable coupler excitations due to the strong always-on capacitive coupling between qubits and coupler.
In contrast, hybridization with the higher excited states of the qubits remains negligible because the system is parked near the zero qubit-qubit coupling condition (inset of Fig.~\ref{fig:sim_sym}(b)).
As the tunable coupler frequency is increased toward the qubit frequencies, the effective qubit-qubit coupling increases and the detunings with states involving coupler excitations decrease.
The $\dressedstate{11,0}$ state hybridizes with the qubit two-excitation $|02,0\rangle$ as the coupler-mediated qubit-qubit interaction grows.
Hybridization with the states $|01,1\rangle$ and $|10,1\rangle$ remains dominant over $|02,0\rangle$ because the qubit-qubit coupling is nearly two orders of magnitude smaller than the qubit-coupler coupling.
The remaining qubit's two-excitation state ($|20,0\rangle$) exhibits negligible hybridization due to its larger detuning from $|11,0\rangle$.

In general, the evolution between two instantaneous eigenstates $|i\rangle$ and $|k\rangle$ is adiabatic if transitions between them remain negligible during the gate, as quantified by the following condition:
\begin{equation}
\begin{aligned}
\left| \frac{\langle i|\dot{k}\rangle}{\omega_k-\omega_i} \right| & =
\hbar\left| \frac{\langle i|dH/d\omega_c|k\rangle}{(E_k-E_i)^2} \right| \left| \frac{d\omega_c}{dt} \right| \\
& = D_{ik}(\omega_c)\left| \frac{d\omega_c}{dt} \right| \ll 1
\label{eq:dfactor}
\end{aligned}
\end{equation}
where $\dot{k}=d|k\rangle/dt$, $\omega_c/2\pi$ is the tunable coupler frequency, and $D_{ik}(\omega_c)$ the adiabatic factor associated with the transition between the  two eigenstates.
In this expression we assume that the Hamiltonian depends only on the tunable coupler frequency.
The adiabatic factor associated with a given state $k$ is defined as the sum of all contributions from states that couple to it, $D_k(\omega_c)=\sum_{i\neq k}D_{ik}(\omega_c)$.
Fig.~\ref{fig:sim_sym}(c) shows the simulated adiabatic factor for the trajectory $\dressedstate{11,0}$ (black solid line), together with the individual contributions $D_{i, \dressedstate{11,0}}$ (colored lines) from each state $i$ in the two-excitation manifold.
The frequency at which each contribution reaches its maximum is indicated in Fig.~\ref{fig:sim_sym}(a) and (b) by colored vertical dashed lines, while the corresponding peak values are marked by solid dots in (c).
The strongest adiabatic constraint arises from the channel involving the $\dressedstate{02,0}$ state.
This is primarily due to its relatively small energy separation from the computational $\dressedstate{11,0}$ state, even though it is not the state with the largest hybridization (Fig.~\ref{fig:sim_sym}(b)).
Interestingly, the contributions associated with the $\dressedstate{20,0}$ and $\dressedstate{01,1}$  states are of comparable magnitude, despite their substantially different hybridization with $\dressedstate{11,0}$.
This highlights that the adiabatic constraint depends not only on the hybridization strength but also on the energy separation between states.

\section{Device set-up and characterization}
\label{sec:setup}
The measured device consists of two fixed-frequency qubits connected by a floating symmetric coupler. We target a qubit-qubit detuning $\Delta_{12}\sim\frac{1}{2}|\eta|/2\pi$ (where $\eta$ is the qubit anharmonicity) in order to balance the following considerations: staying in the straddling regime to ensure an idling point with $ZZ=0$ and at the same time maximizing energy-level separations between $\ket{11}$-$\ket{02}$ and $\ket{10}$-$\ket{01}$ to minimize simultaneous single-qubit errors induced by microwave crosstalk \cite{Kelly2014} and to minimize non-adiabatic errors. Since the device uses fixed-frequency qubits, the targeted qubit-qubit detuning is realized using the alternating-bias assisted annealing technique, which tunes the resistance of Josephson junctions (JJs) at room temperature prior to the device cool-down \cite{Pappas2024,XWang2024}. The qubits on the measured device have a detuning $\Delta_{12}\simeq98$ MHz that is close to half of the qubit anharmonicities ($\eta_1/2\pi=-227$ MHz and $\eta_2/2\pi=-221$ MHz). 

\begin{figure}[tb!]
    \centering
    \includegraphics{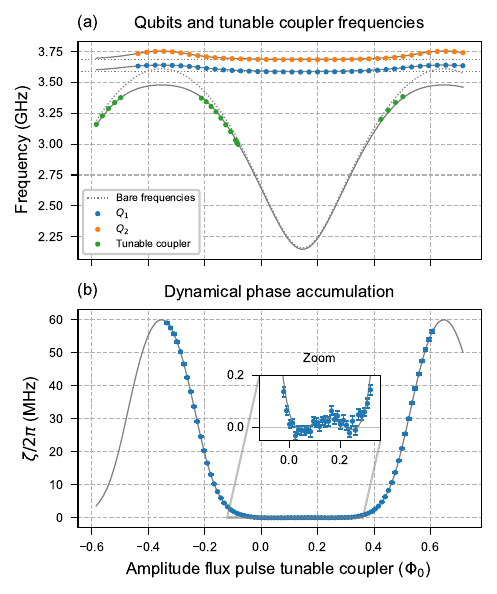}
    \caption{Joint fit of spectroscopic data and dynamical phase accumulation versus amplitude of the flux pulse to the tunable coupler $\Phi_c$. The residuals of $\zeta$ are rescaled by a factor 100; qubit frequencies and anharmonicities are held constant and come from the measured values at zero coupling.
    (a) Measured qubit and tunable coupler frequencies (colored dots) together with the best fit (gray solid lines). Gray dashed lines are the bare frequencies from the fit.
    (b) Experimental data for the dynamical phase accumulation measured with the JAZZ sequence (dots with error bar) and best fit (gray solid line). Inset is a zoom-in near the idling flux. The crossing at $\zeta=0$ is captured both by the measurements and fit.
    }
    \label{fig:freqs_and_zz}
\end{figure}

In the following experiments, the tunable coupler is dc-biased at $\Phi_{c,\mathrm{dc}}=0.35\Phi_0$, where residual $ZZ=27.3$~kHz. Although an exact $ZZ=0$ operating point is accessible on this device (Fig.~\ref{fig:freqs_and_zz}) and is preferable for simultaneous single-qubit gates, we selected a coupler bias with a small non-zero $ZZ$ coupling to improve readout performance during the experiments involving two-qubit gates. At the selected coupler bias, single-qubit gate fidelities measured with randomized benchmarking exceed 99.95\% when the gates are run in isolation and simultaneously. Single-qubit gate times of Q1 and Q2 are 60~ns and 36~ns, respectively. Additional details and characterization of the qubits at this idling point are included in Section~\ref{sec:supplement_experimentsetup} of the supplement.

\subsection{Spectroscopy data and dynamical $\zeta$-rate}
\label{sec:device_characterization}
We characterize the dressed frequencies of the system as a function of flux applied to the superconducting quantum-interference device (SQUID) of the coupler $\Phi_c$ in Fig.~\ref{fig:freqs_and_zz}(a). Qubit frequency data $f_1(\Phi_c)$ and $f_2(\Phi_c)$ are measured by flux-pulsing the coupler during the delay time of a Ramsey experiment. Coupler frequency $f_c(\Phi_c)$ is measured with three-tone spectroscopy by sending a probe tone to the drive line of one of the qubits, performing a Ramsey measurement on the other qubit to detect a qubit-frequency shift when the coupler is excited, and flux-pulsing the coupler during the Ramsey delay time to change its frequency (similar to \cite{Marxer2023}).

To validate that the tunable values for $ZZ$ can be turned off completely and also have a high dynamic range when the qubits are in the straddling regime, we measure how the dynamical $\zeta$ between qubits changes with the amplitude of the coupler flux pulse (Fig.~\ref{fig:freqs_and_zz}(b)). This measurement is done with the joint amplification of $ZZ$ interaction (JAZZ) method \cite{Sagastizabal2021,KuPlourde2020}, in which a Ramsey sequence is performed on a target qubit and $\pi$ pulses are sent to both qubits halfway through the sequence to induce a phase-shift arising from the $ZZ$ interaction; by flux pulsing the TC during the times between the Ramsey $\pi/2$ pulses and the control $\pi$ pulses, we map out how $\zeta$ changes with $f_c$ \cite{GotoNakamura2024}. 

The frequency of the coupler can be tuned to a point of perfect cancellation where $ZZ=0$ (inset of Fig.~\ref{fig:freqs_and_zz}(b)). When $f_c$ moves towards the qubit frequencies, $\zeta$ increases unidirectionally from $\zeta/2\pi=0$ up to 60~MHz. On this device, the maximum $\zeta$ is limited by the maximum frequency of the coupler rather than by restrictions in the two-excitation manifold. Because of the favorable energy level ordering of $\dressedstate{11}$ in the symmetric coupler system, there is an unbounded increase in the conditional phase as coupler frequency increases, so $\zeta$ keeps becoming more positive. In comparison, $\zeta$ for the asymmetric coupler system changes signs and eventually saturates (Section~\ref{sec:supplement_symVasym} for a direct comparison between symmetric and asymmetric coupler systems).

We extract values for the Hamiltonian and TC transmon parameters from a joint fit of the dependence of $f_k$ (where $k=1,2,c$) and $\zeta$ on $\Phi_c$. Due to the small values of $\zeta$ near $ZZ=0$, residuals of $\zeta$ are rescaled by a factor of 100 for the joint fit. The extracted values, shown in Table~\ref{tab:supplement_fitParams} of the supplement, are close to design values and are the parameters we use for the simulations in Fig.~\ref{fig:sim_sym} and Fig.~\ref{fig:leakage}(e--g).  In the fit to the frequencies of the qubits and coupler (dashed line in Fig.~\ref{fig:freqs_and_zz}(a)), we highlight that $\ket{00,1}$ is almost on resonance with $\ket{10,0}$ when the TC is at its maximum frequency. 

\subsection{Adiabatic gate implementation and leakage characterization}
We implement the adiabatic CZ gate by applying a single dc-pulse to the tunable coupler. For each pulse duration and shape, we calibrate the gate by selecting the coupler pulse amplitude that is needed to accumulate conditional phase $\phi_\mathrm{CPHASE}=\pi$ using the JAZZ2-$N$ sequence from Ref.~\cite{GotoNakamura2024}. 
An IIR filter is also applied to the coupler pulse to correct for medium- to long-time constants (in the range 50~ns--10~$\mu$s) coming from memory effects in the pulse shape \cite{Bao2022}. 

As an initial pulse shape, we consider an envelope inspired by the Slepian-pulse from Ref.~\cite{Martinis2014}, which has been used for two-qubit gates in previous works due to its favorable power spectral density profile \cite{Sung2021,Marxer2025,GotoNakamura2024,Glaser2025,VallesSanclemente2026}. The dc-pulse envelope is constructed from a Fourier cosine series: 
\begin{equation}
    V(t) = \sum_{n=1,2,...}a_n\left[1-\cos\left({\frac{2\pi nt}{t_\mathrm{CZ}}}\right)\right],
    \label{eq:fourier_cos}
\end{equation} where $a_n$ is the weight of the $n$-th Fourier cosine term, $t_\mathrm{CZ}$ is the gate time, and the odd $a_n$ terms sum up to 0.5. We simplify tune-up of the CZ gate by defining the pulse envelope---rather than slope of the qubit frequency---in terms of the Fourier series and using the first two Fourier terms (fixing $a_1=0.5$, varying $a_2$).

\begin{figure}[tb]
    \centering
    \includegraphics[width=\columnwidth]{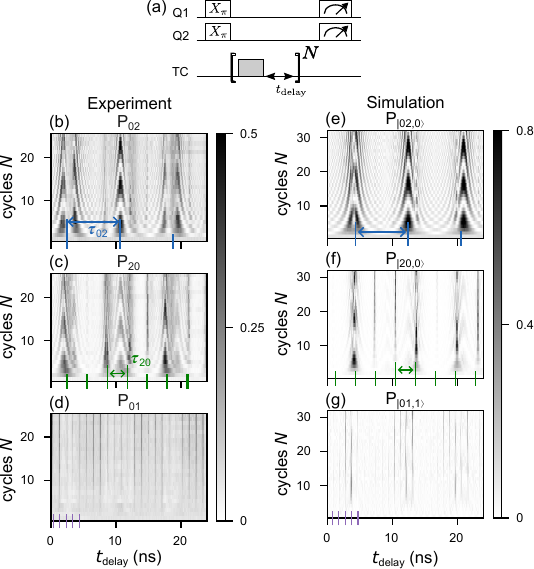}
    \caption{Leakage amplification experiment with a 20~ns cosine pulse for the states with largest adiabatic factors. (a) Pulse sequence used to measure leakage: repeating $N$ cycles of the flux pulse with a varying delay $t_\mathrm{delay}$ between each pulse. (b--d) Experimental data of the projections onto the $\ket{02}$, $\ket{20}$, and $\ket{01}$ states of the qubits using their three-state classifiers. (e--g) QuTiP simulations of the projections onto $\ket{02,0}$, $\ket{20,0}$, and $\ket{01,1}$ using the same pulse shape used in the experiment (parameter details are included in the text). Peak intervals $\tau_L$ corresponding to the idling frequency-detuning between the leakage states $\ket{L}$ and $\ket{11,0}$ are marked with vertical colored lines and line up with most of the peaks in the data and simulation. We can infer that the measured $P_{01}$ peaks seen in (d) come from qubit-coupler population exchange due to the close alignment between the expected idling interval (purple lines) and the measured peak intervals. The $P_{02}$ and $P_{20}$ exhibit qualitatively faster $N$ oscillations than $P_{01}$ in (b--c, e--f) compared with (d, g), which is consistent with stronger involvement of the $\ket{02,0}$ and $\ket{20,0}$ leakage channels than of the $\ket{01,1}$  channel in the population dynamics. This interpretation is further supported by the population dynamics discussed in Section~\ref{sec:supplement_dfactor}. The discrepancy between (b--d) and (e--g) in the oscillations along $N$ likely reflects an experiment-simulation mismatch in the details of the parameters used in the high-coupling regime.}
    \label{fig:leakage}
\end{figure}
We use a leakage-amplification measurement to test whether the observed dominant leakage channels are consistent with the relative adiabatic-factor values of the system. Following Ref.~\cite{Marxer2025}, we prepare the qubits in $\ket{11}$ and apply $N$ cycles of flux pulses. In the presence of non-adiabatic transitions from $\ket{11,0}$ to leakage state $\ket{L}$, repeated application of the CZ gate induces population exchange between the two states, characterized by a swapping angle $\theta$. States $\ket{11,0}$ and $\ket{L}$ also acquire a relative phase difference $\phi$. 

The resulting signal after $N$ cycles can be written in the general form $A+B\cos(2N\mu)$, where $A$ and $B$ depend on $\theta$ and $\phi$ and $\cos\mu=\cos(\phi/2)\cos(\theta/2)$ \cite{Marxer2025}. In the specific case of $\phi=0$, the oscillation frequency along $N$ becomes dependent only on the population leakage term: $2\mu=\theta$. Therefore, under this condition, flux pulses that induce more non-adiabatic transitions will produce faster oscillations as a function of $N$. In our experiment, we cancel out $\phi$ by adding a delay $t_\mathrm{delay}$ after each pulse. Constructive signal amplification occurs periodically as a function of $t_\mathrm{delay}$, with peak spacing intervals of $\tau_L=1 /(f_{\ket{11,0}}-f_{\ket{L}})$, where $f_{\ket{11,0}}-f_{\ket{L}}$ is the detuning of the leakage state from $\ket{11,0}$ at idling. As a result, the leakage state can be identified from the observed periodicity along $t_\mathrm{delay}$, and the oscillation rate along $N$ provides a qualitative indication of the leakage transition strength.

We run the leakage amplification measurement for a 20~ns gate with a cosine envelope ($a_2=0$), applying $N$ cycles of the pulse and varying $t_\mathrm{delay}$ between the pulses as depicted in Fig.~\ref{fig:leakage}(a). The maximum pulse frequency of this 20~ns CZ gate---$f_c\approx3.5$ GHz---is close to the maximum coupler frequency. At this operating point, the eigenstates with the largest $D_{i,\dressedstate{11,0}}$ values are $\dressedstate{20,0}$, $\dressedstate{02,0}$, and $\dressedstate{01,1}$ (Fig.~\ref{fig:sim_sym}(c)). We therefore compare the relative leakage strengths of these channels using the measured populations $P_{20}$, $P_{02}$, and $P_{01}$ obtained by projecting onto the corresponding qubit states ${\ket{20}}$, ${\ket{02}}$, and ${\ket{01}}$ (Fig.~\ref{fig:leakage}(b--d)). For additional validation, we also run QuTiP simulations of $P_{\ket{20,0}}$, $P_{\ket{02,0}}$, and $P_{\ket{01,1}}$ using the same pulse shape (20~ns, $a_2=0$) and using coupler parameters and frequency-independent coupling parameters from the joint-fit results shown in Table~\ref{tab:supplement_fitParams}. The simulation pulse starts at a coupler idling frequency of 2.54~GHz and ends at a bare coupler frequency value of 3.51~GHz. We plot the simulation results in Fig.~\ref{fig:leakage}(e--g). 

We first look at the intervals between signal peaks to identify the oscillations that correspond to the relevant leakage states. This is especially relevant for identifying the leakage channel between $\ket{01,1}$ and $\ket{11,0}$ since we cannot directly read out the tunable coupler. We note that the peak interval $\tau_{01}\approx1$~ns in the $P_{01}$ data is consistent with the 940 MHz detuning between $\ket{01,1}$ and $\ket{11,0}$ (represented by purple lines in Fig.~\ref{fig:leakage}(d,g)), so we can infer that the peaks in $P_{01}$ correspond to leakage to the coupler. 

While we can directly read out the second-excited state of the qubits, the population dynamics in $P_{20}$ and $P_{02}$ exhibit more complicated interference patterns due to strong coupling and higher-order non-adiabatic processes for a 20~ns CZ gate (Section~\ref{sec:supplement_dfactor}). However, the relevant leakage features can still be identified by matching their $t_\mathrm{delay}$ periodicities to the expected idle-state detunings from $\ket{11,0}$, which are marked by the blue and green lines in Fig.~\ref{fig:leakage}(b--c) and (e--f). The oscillation rate along $N$ then provides a qualitative indication of leakage strength. Additional details of the interference structure and comparison with simulation are given in Section~\ref{sec:supplement_dfactor}. 

With these leakage channels identified, we compare their relative strengths with the simulated adiabatic factors in Fig.~\ref{fig:sim_sym}(c). The measured oscillation rates along $N$ are faster in $P_{02}$ and $P_{20}$ than in $P_{01}$, despite the strong hybridization between the $\dressedstate{11,0}$ and $\dressedstate{01,1}$ states near $f_c\approx3.5$ GHz. Instead, the measured oscillation trend is consistent with the relative values of $D_{i,\dressedstate{11,0}}$, which are largest for transitions from $\dressedstate{11,0}$ to $\dressedstate{02,0}$ and $\dressedstate{20,0}$. This agreement supports the conclusion from simulations in Fig.~\ref{fig:sim_sym}(c) that the dominant non-adiabatic transitions for the 20~ns gate involve the qubits’ second excited states $\ket{02,0}$ and $\ket{20,0}$ instead of leakage to the coupler. The leakage amplification data are also consistent with leakage strength being governed by the combination of level separation and hybridization captured by the adiabatic factor, rather than hybridization alone. 

\section{Accessing fast adiabatic CZ gates}
The dynamical $\zeta$-rate on the symmetric coupler system can unlock fast adiabatic CZ gate speeds. To assess performance of the adiabatic CZ gate on this system, we benchmark the gate with various durations and with different pulse shapes. In the following sections, we calibrate the gate pulse for a CZ gate and correct for single-qubit phase shifts with a virtual-Z phase correction (measured with the robust phase estimation technique \cite{Kimmel2015,Russo2021}) on the single-qubit gates. 

Using standard randomized benchmarking (RB) and interleaved randomized benchmarking (iRB) experiments \cite{MagesanIRB2012,MagesanRB2012,Corcoles2013}, we benchmark the CZ gates by fitting RB and iRB decay curves with maximum likelihood estimation (MLE), where we use the 95\% confidence interval for the sequence fidelities. We report the average RB error per gate $r_\mathrm{RB}$ and CZ error per gate $r_\mathrm{CZ}$, where $r_{\mathrm{CZ}}=\frac{d-1}{d}\left(1-\frac{p_{\mathrm{iRB}}}{p_{\mathrm{RB}}} \right)$, $d$ is the Hilbert space dimension, and $p_{\mathrm{RB}}$ and $p_\mathrm{iRB}$ are the depolarizing factors of the RB and iRB experiments, respectively. Details of the MLE fitting procedure, confidence interval estimation, and gate-error calculations are provided in Sections~\ref{sec:CI_binomial}--\ref{sec:supplement_montecarlo}. We note that leakage is non-negligible at the shortest gate times and can complicate the interpretation of standard RB and iRB decay parameters \cite{Epstein2014,Hashim2025,YHChen2025}. We nevertheless use the standard RB-derived errors to characterize the gate-performance trends across pulse parameters and separately quantify leakage for an optimized 24~ns gate below. 

\subsection{Minimal pulse shaping}
\begin{figure}[!tb]
    \centering
    \includegraphics[width=\columnwidth]{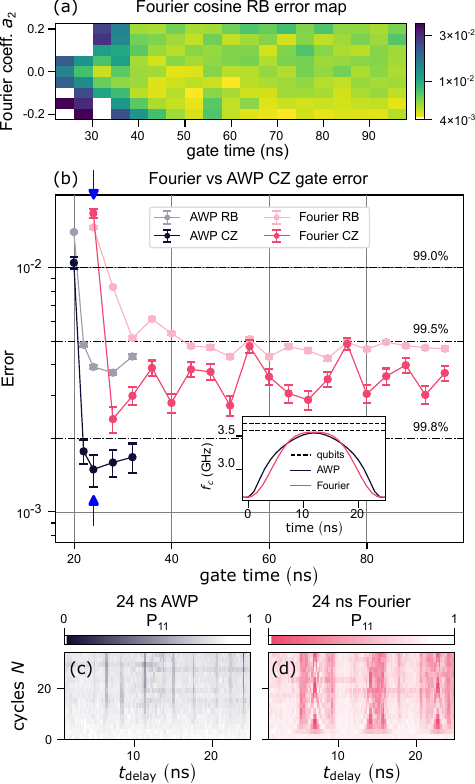}
    \caption{Adiabatic CZ gate with different pulse shapes. (a) RB error for gates with Fourier cosine shape, varying $a_2$ at each gate time. When $t_\mathrm{CZ}>44$ ns, RB error variation is small across the range of scanned $a_2$ values. The range of $a_2$ values with low error is more restricted at shorter gate times. (b) Comparing errors of gates enacted by the Fourier pulse with the best $a_2$ coefficient (magenta) and by the AWP (dark blue). AWP provides a boost to gate fidelity relative to Fourier pulse at fast gate times, maintaining CZ (RB) fidelities above 99.8\% (99.5\%) down to 22 ns. In the inset, we compare the two pulse shapes used to enact a 24~ns gate. Dashed lines correspond to the idling frequencies of the qubits. Blue arrows in the main plot indicate the data points that correspond to 24~ns. (c) Leakage out of $\ket{11}$ for a 24~ns AWP-shaped gate. Oscillations in $N$ are lower for the AWP-shaped gate than for a 24~ns Fourier cosine pulse with the same maximum $f_c$, shown in (d), which indicates that AWP can reduce leakage at fast gate times.}
    \label{fig:fourierVawp}
\end{figure}
By using the two-term Fourier cosine pulse described in Eq.~\ref{eq:fourier_cos}, the tune-up procedure of the adiabatic CZ gate is simplified to the calibration of two parameters: at each gate time $t_\mathrm{CZ}$, we select a value of the Fourier coefficient $a_2$ to define the pulse envelope and find the TC flux amplitude using JAZZ2-$N$. 

In Fig.~\ref{fig:fourierVawp}(a), we plot the benchmarking results from the tune-up procedure. The amount of RB error variation across $a_2$ at each gate time reflects the sensitivity of gate performance to the pulse shape. For example, for CZ gate times longer than 44\,ns, RB fidelity varies by a small amount across the range of scanned $a_2$ coefficients ($\approx$99.2--99.6\% for $a_2=-0.2$ to $0.2$), which suggests that the gates are not sensitive to the specific trajectory of the pulse if the gate time is long.  There is a slight fidelity improvement for pulse shapes with $a_2<0$ that is likely coming from an incoherent error improvement: when $a_2<0$ values, pulses have a slower rise time, so the coupler spends more time near its idling frequency where the qubit coherence times are higher (Section~\ref{sec:supplement_cohFlux}) when compared with pulses with $a_2>0$. 

When the gate time becomes shorter, the coupler frequency needs to be pushed to higher values where it becomes more difficult to maintain adiabaticity and suppress transitions to non-computational states. As a result, as the gate time drops below 44\,ns, the fidelity measured using standard RB not only becomes more sensitive to $a_2$ but also begins to decrease. In this gate time range, the maximum pulse frequency exceeds 3.4\,GHz, which corresponds to the point where the coupler frequency approaches the maximum $D_{i,\dressedstate{11}}$ value in Fig.~\ref{fig:sim_sym}(c).

In Fig.~\ref{fig:fourierVawp}(b), we plot the RB error and the CZ error extracted from iRB at each gate time after selecting the Fourier coefficient $a_2$ that gives the lowest error (magenta points). Based on these extracted points, it is evident that an adiabatic CZ gate with high fidelity is possible with minimal pulse shaping (varying a single coefficient of a Fourier cosine envelope): RB fidelity drops below 99.5\% starting at a gate time of 40~ns and below 99\% for a gate time of 24~ns.

\subsection{Edge-tailored pulse shaping for faster gates}
Below 40~ns, leakage begins to increase, so we prototype an experimental procedure to reduce non-adiabatic transitions and therefore make it possible to take advantage of the fast gate times that are accessible in the symmetric coupler system. In particular, we generate the adiabatically weighted pulse (AWP) that was proposed in Ref.~\cite{FeiYan2021} and is implemented by weighting the speed of the pulse with the instantaneous total $D$-factor: 
\begin{equation}
\begin{aligned}
D(\omega_c)= \sum_{k\in K}D_k(\omega_c)\\
\frac{\mathrm{d}\omega_c}{\mathrm{d}t}=\frac{\lambda}{D(\omega_c)}\sin\left(\frac{2\pi t}{t_\mathrm{CZ}}\right), \label{eq:awp}
\end{aligned}
\end{equation}
where $D_k$ from Eq.~\ref{eq:dfactor} is summed over all the computational states $K$ of the system and $\lambda$ is the AWP coefficient. With this pulse shaping, the gate evolves more slowly when non-adiabatic transitions (as quantified by $D$) are more likely. 

We tune-up the AWP pulse by (1) estimating $D(\omega_c)$ from measured qubit characteristics and design values; (2) converting the frequency pulse envelope generated in Eq.~\ref{eq:awp} to flux amplitude using parameters extracted from a fit to coupler spectroscopy data; and (3) using the pulse sequence from JAZZ2-$N$ to select the $\lambda$-coefficient for which the pulse accumulates $\phi_\mathrm{CPHASE}=\pi$ (Section~\ref{sec:supplement_awptuneup} in the supplement). In order to mitigate flux line non-idealities and fine-tune the final pulse shape, we also optimize over 3 parameters of the Hamiltonian ($g_{12}$, $g_{1c}$, $g_{2c}$) and 3 parameters that define the $f_c$ trajectory (the product of the Josephson and charging energies $E_JE_C$, the ratio of the JJs, and the coupler anharmonicity) using the hyperparameter optimization software Optuna \cite{optuna2019}. The cost function is  $\log(\mathrm{CZ\,error})$ for all gate times except $t_\mathrm{CZ}=20$ ns (which uses $\log(\mathrm{RB\,error})$ due to the high amount of error for this gate pulse). See Section~\ref{sec:supplement_awptuneup} in the supplement for the pulse shapes output by the Optuna optimizer. 

We plot the fidelities of the AWP CZ gate (dark blue) and compare with the Fourier cosine CZ gate (magenta) in Fig.~\ref{fig:fourierVawp}(b). For fast gate times, the AWP provides a boost to 2Q gate performance over the gates enacted with the Fourier cosine pulse. In particular, while the RB fidelity of the Fourier cosine CZ gate drops below 99.5\% for 40~ns and shorter, the RB fidelity with AWP stays above 99.5\% at 24 ns (blue arrows in Fig.~\ref{fig:fourierVawp}(b)), unlocking an additional 16~ns of gate speed based on standard RB fidelities. Furthermore, the CZ fidelity with AWP is higher than CZ fidelity with Fourier cosine pulse for all of the calibrated gate times. 

\begin{figure}[tb!]
    \centering
    \includegraphics[width=\columnwidth]{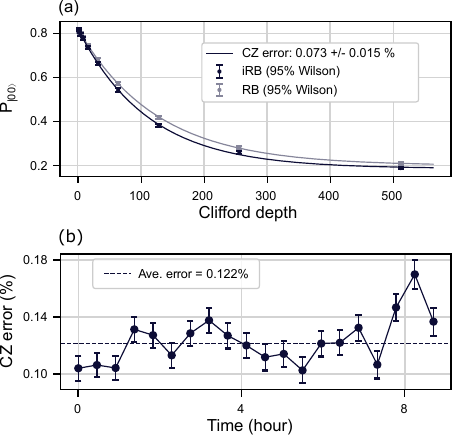}
    \caption{Benchmarking a 24~ns adiabatic CZ gate. (a) Randomized benchmarking curve with CZ fidelity of 99.927$\pm$0.015\%. RB experiment points are in the 95\% Wilson confidence intervals (Section~\ref{sec:CI_binomial}). (b) An 8.5-hour time trace of the gate implemented in (a) yields an average CZ fidelity of 99.878$\pm$0.016\%.}
    \label{fig:benchmarking}
\end{figure}
Therefore, even though the tune-up complexity increases with the AWP, incorporating details of the energy levels into the pulse-shaping process provides an improvement to fidelities at fast gate times. A comparison between the two types of pulse shapes---defined by a single generic parameter versus multiple Hamiltonian parameters---is shown in the inset of Fig.~\ref{fig:fourierVawp}(b) for a 24~ns CZ gate. At this gate speed, the AWP has a smaller frequency slope than the Fourier cosine pulse as the coupler reaches the maximum $f_c$ of the pulse, which is within 150~MHz of the dressed qubit frequencies. 

Monitoring the $\ket{11}$ state with the leakage amplification measurement also more directly illustrates the impact of the pulse shapes on non-adiabatic transitions. In Fig.~\ref{fig:fourierVawp}(c,d), we compare the leakage dynamics due to the 24~ns AWP CZ gate (which is the gate time at which Fourier RB fidelity drops below 99\%) with a 24~ns Fourier cosine pulse that has the same maximum coupler frequency. Here, we highlight that the leakage landscape induced by the AWP is much simpler and has slower oscillations along $N$ than the one induced by the unshaped cosine pulse, indicating that the AWP does indeed reduce undesirable leakage out of $\ket{11}$.  This difference in leakage dynamics between Fig.~\ref{fig:fourierVawp}(c) and (d) demonstrates how a pulse shape tailored to the edge parameters can help improve performance at fast gate speeds even when maintaining adiabaticity becomes more challenging in the high $D$-factor regime. 

Finally, we test the stability of a 24~ns adiabatic CZ gate, with 2-ns of wait time before and after the active flux duration. With an additional brute-force scan starting from the parameters obtained from an initial Optuna search, we reach a CZ fidelity of 99.927$\pm$0.015\% as shown in Fig. \ref{fig:benchmarking}(a), demonstrating that it is possible for a 24~ns adiabatic CZ gate to reach a high fidelity by weighting the pulse with the adiabatic $D$-factor. At a later time, we benchmark this gate over 8.5 hours, which yields a time-averaged CZ fidelity of 99.878\% with a standard deviation of 0.016\% (Fig.~\ref{fig:benchmarking}(b)). The gate-fidelity of the 24~ns AWP-shaped adiabatic gate is close to the coherence-limited fidelity imposed by the qubit coherence with tunable coupler flux modulation (Section~\ref{sec:supplement_cohFlux}), and the increase in CZ error near the end of the time trace comes from a decrease in qubit coherence times (Section~\ref{sec:supplement_cohIdling} in the supplement). We note that these fidelities are extracted using a standard RB experiment, which does not explicitly account for leakage. As a complementary metric, we implement a separate leakage RB measurement \cite{Wood2018,Rol2019,Sung2021} using the same pulse parameters, which gives a leakage rate of $0.027\% \pm 0.01\%$ (Section~\ref{sec:supplement_leakage}). A more in-depth error-budgeting will be the focus of future work that will investigate the fidelity impact coming from qubit coherence times with TC flux-modulation, coherence times in the two-excitation manifold, and non-adiabatic transitions. 

While our studies have been focused on a device with two qubits and a coupler, the adiabatic gate implemented on the symmetric coupler architecture also performs well in larger-scale systems. As a demonstration at larger scale, we present data taken on 36-qubit devices composed of four 9-qubit chiplets in Section~\ref{sec:supplement_36q}. This set of data demonstrates that (1) the adiabatic gate achieves performance comparable to or better than that of diabatic parametric-resonance gates \cite{Sete2021prgates} and (2) with minimal pulse shaping, a median CZ fidelity of 99.48\% can be achieved across 55 qubit-qubit edges, with gate times ranging from 40~ns to 144~ns. Thus, while the detailed characterization in this work is performed on a two-qubit system, the same adiabatic gate approach on a symmetric coupler architecture also performs well using only minimal pulse shaping across a larger, chiplet-based device with higher connectivity. 

In summary, the favorable energy-level ordering of the symmetric coupler system enables fast, high-fidelity two-qubit gates while keeping the qubits in the straddling regime, thereby preserving single-qubit gate performance and maintaining a $ZZ=0$ idling condition. Simulations reveal unbounded growth of the dynamical $\zeta$-rate with only a gradual increase in leakage, as quantified by the adiabatic factor $D$. 

We validate these findings experimentally in a simple and scalable system consisting of fixed-frequency qubits and a tunable coupler with typical transmon parameters. The CZ gate is activated by a single flux pulse on the coupler, thereby avoiding tunable qubits and associated flux noise. Due to the favorable energy structure of the symmetric coupler system, we achieve high-fidelity gates down to 44~ns without aggressive pulse-shaping. To access even shorter gate times, we introduce the adiabatic factor $D$ to the pulse-shaping, which reduces leakage even when maintaining adiabaticity becomes more challenging. Using this approach, we improve the performance relative to the simple Fourier pulse shape at fast gate times where leakage-related errors become increasingly relevant, reaching a CZ fidelity above $\sim\!$99.8\% down to 22~ns and up to 99.927$\pm$0.015\% at 24~ns (with 2~ns of padding before and after the pulse). These results demonstrate that fast, high-fidelity adiabatic CZ gates with interaction-free idling can be realized without added hardware complexity. Moreover, because the floating tunable coupler does not rely on direct qubit-qubit coupling, this gate can be implemented between qubits on separate dies, lifting constraints on scalability and enabling its implementation on chiplet-based 36-qubit devices. Improvements in qubit and coupler coherence times combined with more advanced pulse optimization techniques can further enhance fidelity on this all-transmon, symmetric coupler platform. 

\section{Acknowledgments and author contributions}
We thank Eyob A. Sete for critically reading the manuscript, and Riccardo Cantone, Joseph Valery, Akel Hashim, and Jacob Bryon for fruitful discussions. We are grateful to Benjamin Wood, Marius Hegedus, Daniel Cox, and Benjamin Griffiths for collecting experimental data on the 36-qubit devices. Finally, we thank the entire Rigetti team for their general support and for building the infrastructure that made this work possible.

S.P. conceptualized the project, coordinated efforts, and performed the simulations and analysis in this work. A.Q.C designed and executed the experiments and data analysis. A.Q.C. and S.P. wrote the manuscript and prototyped the adiabatically weighted pulse experimental method together. X.W. participated in critical discussions and provided support on simulation work. S.S. integrated key parts of the experimental methods into the software. All authors contributed to the manuscript. 

%% file: supplementary.tex
\section{Experimental setup and additional device characterization}
\label{sec:supplement_experimentsetup}

\begin{figure}[bt]
    \centering
    \includegraphics[width=\columnwidth]{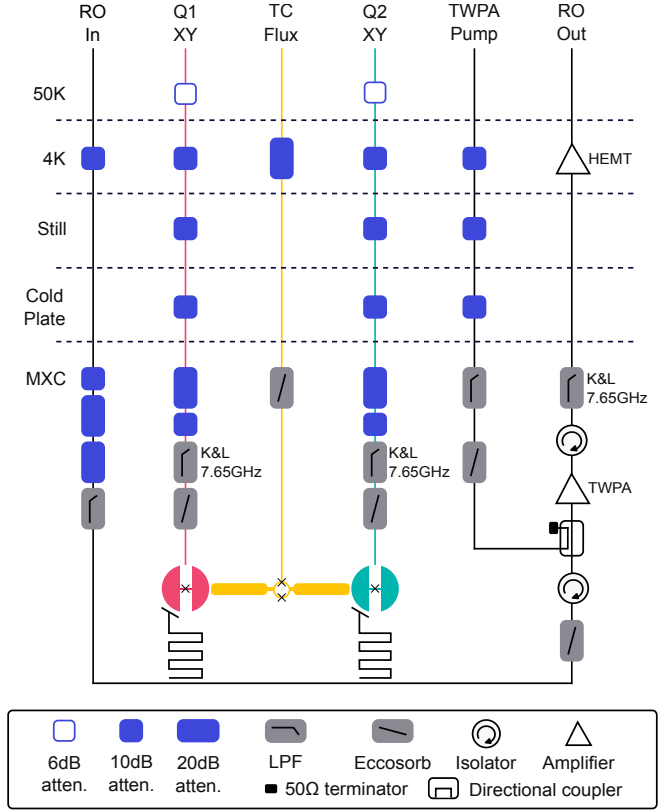}
    \caption{Schematic of the device wiring. Fixed-frequency qubits are coupled via a flux-tunable, symmetric coupler controlled by a dedicated flux bias line. Each qubit has an independent XY microwave drive-line for single-qubit control and is dispersively coupled to a readout resonator. The readout resonators are frequency-multiplexed and coupled to a common transmission line.}
    \label{fig:wiring_diagram}
\end{figure}

The device consists of a chip with the quantum elements (fixed-frequency qubits and tunable coupler) and a cap with signal routing and control wiring. The chips are made using standard fabrication techniques and are flip-chip bonded following the methods described in \cite{Gold2021,Field2024}. Measurements are performed in a dilution refrigerator at a base temperature of approximately 15$\,$mK.  Input lines are attenuated and filtered at each temperature stage to suppress thermal noise. Output signals are routed through cryogenic isolation stages and amplified before room-temperature detection. The sample is shielded by tin-coated aluminum coated with black-absorber and two $\mu$-metal shields (from inside to the outside). Details of the fridge wiring are shown in Fig.~\ref{fig:wiring_diagram}.

There are two $ZZ = 0$ operating points on the measured device at $\Phi_{c,\mathrm{dc}}\sim0.38,\,0.42\Phi_0$ (inset of Fig.~\ref{fig:freqs_and_zz}). Selecting the $ZZ = 0$ point closer to $\Phi_{c,\mathrm{dc}}=0$ would be desirable to minimize the flux delivered to the coupler SQUID and to maintain lower thermal population on the coupler. However, for the two-qubit gate measurements presented in the main text, an idling bias $\Phi_{c,\mathrm{dc}}=0.35\Phi_0$ is applied to the symmetric coupler. At this bias, simultaneous 1Q gate fidelity is greater than 99.9\%, and residual $ZZ=27.3$~kHz, which is slightly off of $ZZ=0$. While $ZZ=0$ is the preferable point of operation for single-qubit gates and is accessible on this device, we moved the coupler due to limitations in the simultaneous readout fidelity that we observed at the time the two-qubit gate benchmarking data was taken. Simultaneous readout performance at $ZZ=0$ recovered after operating points shifted following a thermal cycle, supporting the interpretation that the earlier readout limitation was specific to that experimental operating configuration rather than intrinsic to the $ZZ=0$ condition.  We present additional qubit characteristics measured at the selected idling bias in Table~\ref{tab:supplement_idlingParams}.

\begin{table}[b!]
\centering
\begin{tabular}{lcc}
    \toprule
    & Q1 & Q2  \\
    \midrule
    Frequency (MHz) & 3588 & 3686  \\
    Anharmonicity (MHz) & -227 & -221 \\
    Isolated 1Q fidelity (\%) & 99.96 & 99.96 \\
    Simult. 1Q fidelity (\%) & 99.95 & 99.96 \\
    Readout fidelity (\%) & 85.0 & 94.0 \\
    \bottomrule
\end{tabular}
\caption{Qubit characteristics measured with an idling dc-bias applied to the coupler $\Phi_c=0.35\Phi_0$.}
\label{tab:supplement_idlingParams}
\end{table}

\begin{table}[b!]
\centering
\begin{tabular}{lc}
    \toprule
    TC max. bare frequency (MHz) & 3622\\
    TC idling bare frequency (MHz) & 2644\\
    TC bare anharmonicity (MHz) & -178\\
    TC measured anharmonicity (MHz) & -112\\
    TC JJ ratio & 2.23\\
    $g_{12}$ (MHz) & 3.96\\
    $g_{1c}$, $g_{2c}$ (MHz) & 96.2, 83.9\\
    $\rho_{12}$ & $1.088149\times10^{-3}$\\ 
    $\rho_{1c}$ & $2.669285\times10^{-2}$\\ 
    $\rho_{2c}$ & $2.295814\times10^{-2}$\\ 
    \bottomrule
\end{tabular}
\caption{Parameters extracted from a joint-fit of the dependence of $f_1$, $f_2$, and $\zeta$ on $\Phi_c$  shown in Fig.~\ref{fig:freqs_and_zz}, and the measured dressed anharmonicity of the tunable coupler. The frequency-dependent $g_{ij}$ values are calculated at the measured qubit frequencies and the maximum bare frequency of the coupler.}
\label{tab:supplement_fitParams}
\end{table}
We extract additional parameters related to the tunable coupler and Hamiltonian by performing a joint fit on the data in Fig.~\ref{fig:freqs_and_zz} of the main text, and we present these parameters in Table~\ref{tab:supplement_fitParams}. The bare $g$-coupling values depend on the frequency of the transmons (Section~\ref{sec:g2rho}), so we calculate $g_{ij}$ at the measured qubit frequencies and the maximum bare frequency of the coupler extracted from the fit (Table~\ref{tab:supplement_idlingParams}--\ref{tab:supplement_fitParams}). We also include the frequency-independent coupling parameter $\rho_{ij}$. The dressed coupler anharmonicity comes from a spectroscopy measurement measured at $f_c=3.3\,$GHz and is provided as a comparison to the bare coupler anharmonicity extracted from the fit. 

\section{Simulations and interpretation}
In this section we present QuTiP simulations to support the main claims of the manuscript, and to provide a deeper understanding of the adiabatic mechanism and the identification of the major non-adiabatic channels.
We start by showing how the coupling strengths between transmons (qubits and couplers) is a frequency-dependent parameter and how, under realistic approximations, it can be written in terms of a frequency-independent parameter.
In QuTiP simulations, we work with the frequency-independent coupling strengths to better represent the system.

We will move to a time-dynamic simulation of the leakage amplification experiment, showing how the amplified leakage channels are linked to the adiabatic factors.
With this method, we show how it is possible to predict the main non-adiabatic channels, and we present the impact of a second-order non-adiabatic evolution on the detectable leakage.

Finally, we will compare the symmetric floating coupler layout with the asymmetric configuration.
This final simulation proves the advantage of the symmetric tunable coupler architecture for adiabatic operations.

\subsection{Frequency-independent coupling parameter}
\label{sec:g2rho}
Here we show how the transmon-transmon $g$ interaction strength depends on frequency, and how it is linked to a frequency-independent coupling parameter $\rho$.
The coupling $g_{12}$ between transmon 1 and 2, which can be identified as qubit or coupler, is
$$
g_{12} = \frac{E_{12}}{\sqrt{2}}\left(\frac{E_{J_1}}{E_{C_1}}\frac{E_{J_2}}{E_{C_2}} \right)^{1/4}
$$
with $g_{12}$ and $E_{12}$ the coupling factor and coupling energy, $E_{J_x}$ and $E_{C_x}$ the Josephson and charging energies of transmon $x$~\cite{Sete2021}.
The Josephson energy in a tunable transmon reads:
$$
E_{J_x}(\Phi) = \sqrt{E_{J^1_x}^2 + E_{J^2_x}^2 + 2E_{J^1_x}E_{J^2_x}\cos\left(\frac{2\pi\Phi}{\Phi_0} \right)}
$$
with $E_{J^{1(2)}_x}$ the Josephson energy of junction $1$ ($2$) of qubit $x$.

The frequency of a transmon qubit as a function of flux can be written as
\begin{align*}
    \omega_x(\Phi) &= \sqrt{8E_{J_x}(\Phi)E_{C_x}} - E_{C_x}\left(1+\frac{\xi}{4} \right) \\
    &\simeq \sqrt{8E_{J_x}(\Phi)E_{C_x}} - E_{C_x} \\
    &\simeq \sqrt{8E_{J_x}(\Phi)E_{C_x}}
\end{align*}
since $\xi = \sqrt{\frac{2E_{C_x}}{E_{J_x}}}\ll 1$ and $E_{C_x}$ is much smaller than the typical qubit frequency.
From this last equation we can write the Josephson energy as $E_{J_x}(\Phi) \simeq \frac{\omega_x^2}{8E_{C_x}}$.

The transmon-transmon coupling $g$ can be written as a function of the two transmon frequencies combining the equations above
\begin{align*}
    g_{12} &= \frac{E_{12}}{\sqrt{2}}\left(\frac{E_{J_1}}{E_{C_1}}\frac{E_{J_2}}{E_{C_2}} \right)^{1/4} \\
    &\simeq \frac{E_{12}}{\sqrt{2}}\left(\frac{\omega_1^2}{8E_{C_1}^2}\frac{\omega_2^2}{8E_{C_2}^2} \right)^{1/4} \\
    &= \frac{E_{12}}{\sqrt{2}(64E_{C_1}^2E_{C_2}^2)^{1/4}}\left( \omega_1^2 \omega_2^2 \right)^{1/4} \\
    &=\rho_{12}\sqrt{\omega_1\omega_2}
\end{align*}
with $\rho_{12}$ the frequency-independent coupling term.

\begin{figure}[tb!]
    \centering
    \includegraphics[width=\columnwidth]{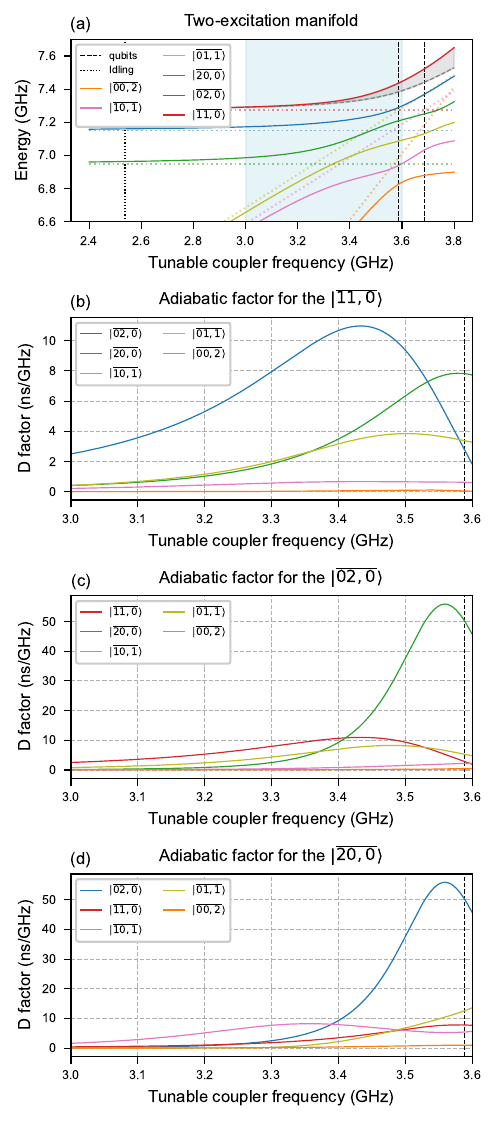}
    \caption{Simulations of multiple adiabatic factors to identify major leakage channels during adiabatic evolution.
    (a) Energy profile in the two-excitation manifold. Gray area denotes the dynamical phase accumulation in adiabatic trajectory, light blue area identifies the span in tunable coupler frequency for the simulation of the adiabatic factors and the target frequencies span for the simulation of the leakage amplification experiments.
    (b), (c), and (d) Adiabatic factors $D_{i,\dressedstate{11,0}}$, $D_{i,\dressedstate{02,0}}$ and $D_{i,\dressedstate{20,0}}$ respectively.
    }
    \label{fig:D_second_order}
\end{figure}
\subsection{Adiabatic factor and non-adiabatic channels}
\label{sec:supplement_dfactor}
In this section, we look more closely at how the adiabatic factor relates to leakage dynamics in simulations. The energy levels of the two-excitation manifold from the main text are replotted in Figure~\ref{fig:D_second_order}(a) for reference, and we focus specifically on the range 3--3.6\,GHz (light blue area in Figure~\ref{fig:D_second_order}(a)). For this frequency range, we plot the two-excitation adiabatic factors $D_{ik}$ that quantify the non-adiabatic transition from state $k$ to leakage state $i$ in Figure~\ref{fig:D_second_order}(b--d) and compare them with simulations of the leakage amplification experiment described in Fig.~\ref{fig:leakage}(a) of the main text.

\begin{figure}[tbp]
    \centering
    \includegraphics[width=\columnwidth]{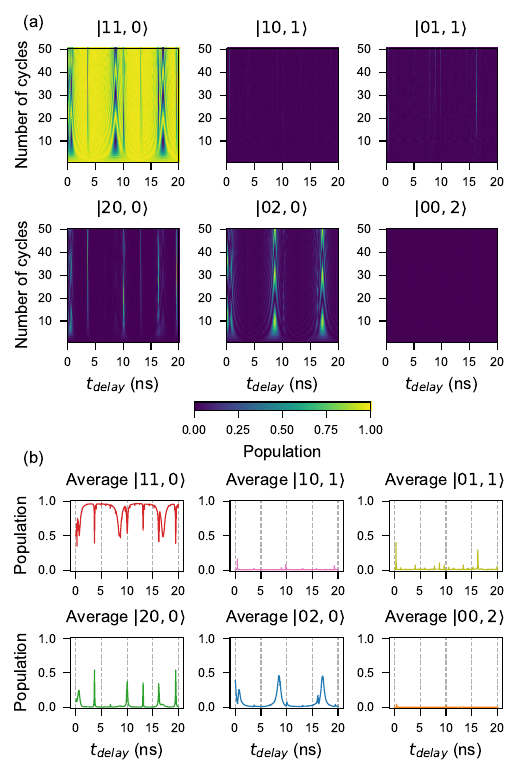}
    \caption{Simulation of the leakage amplification sequence. The system is initially prepared in the $\ket{11}$ state, and a sequence of 24~ns pure-cosine pulses are applied to the tunable coupler, with separation $t_{\mathrm{delay}}$ between pulses. The tunable coupler frequency is modulated from idling at 2.54~GHz to 3.50~GHz.
    (a) Two-dimensional plot of population on each two-excitation manifold state vs $t_\mathrm{delay}$ and number of pulses applied.
    (b) One-dimensional plot of the population averaged over the number of cycles. This helps identify the major leakage channels at the target modulation frequency of the tunable coupler.}
    \label{fig:leakage_ampl_3500}
\end{figure}

\begin{figure*}[tb]
    \centering
    \includegraphics{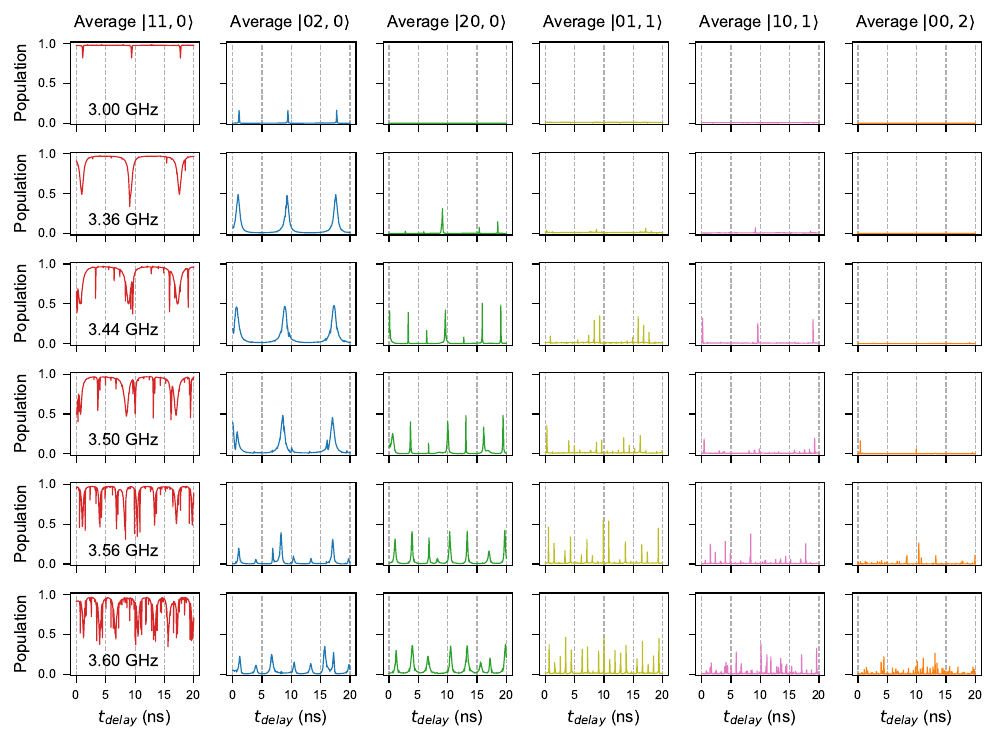}
    \caption{Population averaged over flux pulse cycles. Columns indicate the measured state, where the initial state $\ket{11,0}$ is plotted in the first column. Rows are individual simulations for different amplitudes of the tunable coupler flux pulse. The tunable coupler is modulated from its idling point to the frequency reported in the plot on the first column of each row. These amplitudes are inside the light blue area in Fig.~\ref{fig:D_second_order}(a), and can be mapped with the adiabatic factors reported in Fig.~\ref{fig:D_second_order}(b)-(d). Populations are colored with the same color-scheme used in the main text and in Fig.~\ref{fig:D_second_order}.}
    \label{fig:avgP_vs_target_fc}
\end{figure*}

We simulate the leakage amplification experiment by modulating the tunable coupler frequency from idling to a target value $f_\mathrm{target}$, and we simulate the population on each state while varying the number of pulses and the free-evolution time between pulses. We use a 24~ns Fourier-cosine envelope with $a_2=0$ (pure cosine) for the pulse shape, and we use device parameters derived from the joint fit that are reported in Table~\ref{tab:supplement_fitParams} in the QuTiP simulations. 
This procedure generates two-dimensional plots similar to Fig.~\ref{fig:leakage_ampl_3500}(a).
By averaging the population over the number of cycles, we can collapse the data into a one-dimensional plot to simplify the visualization (Fig.~\ref{fig:leakage_ampl_3500}(b)).
While this one-dimensional plot does not capture all of the dynamics, it provides a simple method to identify leakage channels.
We repeat this leakage simulation for tunable coupler target frequencies from 3.0\,GHz to 3.6\,GHz in intervals of 20\,MHz to look at how the leakage dynamics change as the target frequency moves to increasingly non-adiabatic regimes (increasing values of select frequencies are shown from the top to bottom row of Fig.~\ref{fig:avgP_vs_target_fc}).

The system is initially prepared in the $\ket{11,0}$ state.
Starting in the low coupler-frequency regime,  
we look at the average populations at the end of the train of flux pulses when $f_\mathrm{target}=3.00$\,GHz, which are plotted in the first row of Fig.~\ref{fig:avgP_vs_target_fc}. 
The leakage amplification simulation exhibits a drop in the population of the initial state $\ket{11,0}$ as it bleeds into $\ket{02,0}$.
The time separation between peaks is 8.32\,ns, in perfect agreement with the expected 8.33\,ns from the energy difference $E_{11,0} - E_{02,0}$ at idling from the simulation. Furthermore, we highlight how the non-adiabatic error is largely due to the dynamical bleeding from $\dressedstate{11,0}$ into the $\dressedstate{02,0}$ state. This dominating bleeding channel is well-captured by the relative magnitudes of the $D_{i, \dressedstate{11,0}}$ terms in the low coupler frequency regime ($f_\mathrm{target}\lesssim3.5$~GHz), where $\dressedstate{02,0}$ is the largest term (blue line in Fig.~\ref{fig:D_second_order}(b)). 

When the tunable coupler is modulated up to 3.36~GHz, the non-adiabatic factors are in a moderately large regime: the main non-adiabatic transition $\dressedstate{11,0}\rightarrow\dressedstate{02,0}$ continues to increase, as captured by both the increase of $D_{\dressedstate{02,0},\dressedstate{11,0}}$ and the larger and broader peaks of the leakage amplification simulation (Fig.~\ref{fig:D_second_order}(b), second row of Fig.~\ref{fig:avgP_vs_target_fc} respectively). Furthermore, the adiabatic factors for the transitions from $\dressedstate{11,0}$ to states $\dressedstate{20,0}$ and $\dressedstate{01,1}$ also start becoming non-negligible when $f_\mathrm{target}=3.36$~GHz (green and olive lines, respectively, in Fig.~\ref{fig:D_second_order}(b)), which translates to the emergence of new peaks in the corresponding leakage plots of the second row of Fig.~\ref{fig:avgP_vs_target_fc}. This provides validation that the $D$-factor is a good proxy for leakage dynamics.

Finally, we look at the peak distribution in the large $D$-factor regime, where all the peaks have emerged for a given state. We note that the peak distribution is non-uniform and will explain the origin of this non-uniformity later in the section.  
For now, we want to point out that the separation between peaks matches detuning between energy-levels at idling, which we verify by extracting peak separations at $f_\mathrm{target}$ values after all the peaks have emerged; we have 3.16~ns for $\ket{20,0}$ ($f_\mathrm{target}= 3.44$~GHz) and 0.94~ns for $\ket{01,1}$ ($f_\mathrm{target}= 3.60$~GHz), which perfectly agree with the simulated idling energy gaps of 3.16~ns and 0.94~ns, respectively.
The dynamics of the leakage to $\dressedstate{10,1}$ is more complicated and makes it more difficult to find alignment between peak spacings and energy gap detuning, which we will address at the end of this section.

So far, we have discussed the impact of the non-adiabatic bleeding out of the initial $\ket{11,0}$ state, which we call first-order effects. However, as this leakage accumulates, primarily in the $\ket{02,0}$ and $\ket{20,0}$ states, an additional second-order effect begins to emerge in the leakage amplification simulations, where leakage originates from a state that is not the initially prepared $\ket{11,0}$. The second-order non-adiabatic transitions from $\ket{02,0}$ and $\ket{20,0}$ to other states are captured by the adiabatic factors $D_{i, \dressedstate{02,0}}$ and $D_{i, \dressedstate{20,0}}$ in Fig.~\ref{fig:D_second_order}(c),(d), respectively, and follow dynamics similar to what we have described for the first-order bleeding. 

\begin{figure}[tb]
    \centering
    \includegraphics[width=\columnwidth]{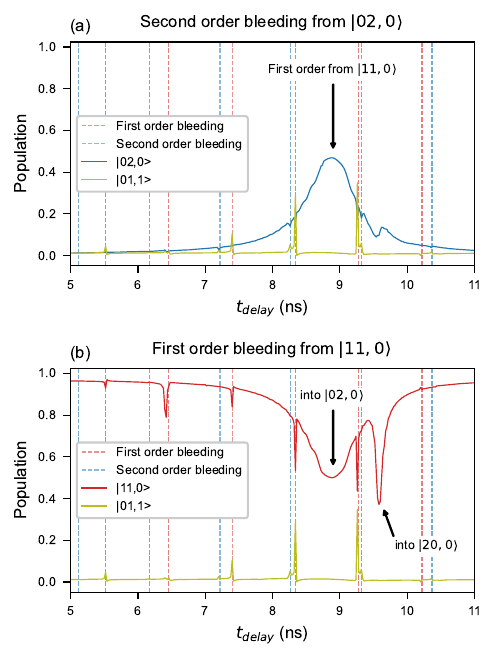}
    \caption{Leakage amplification simulation for a target tunable coupler frequency of 3.44~GHz. Blue and red dashed vertical lines are separated 1.05~ns and 0.94~ns respectively, corresponding to the expected time separation of the second order bleeding $\dressedstate{02,0} \leftrightarrow \dressedstate{01,1}$, and first order $\dressedstate{11,0} \leftrightarrow \dressedstate{01,1}$.
    (a) Matching second-order bleeding peaks on $\ket{01,1}$ with dips in $\ket{11,0}$.
    (b) Matching first-order bleeding peaks and dips. Second-order bleeding peaks, aligned with blue dashed vertical lines, do not show up as dips in the $\ket{11,0}$ trace.}
    \label{fig:1st2nd_order_leakage}
\end{figure}

To illustrate the behavior of the second-order non-adiabatic bleeding in more detail, we focus on a peak that comes from first-order leakage. In particular, we look at $\ket{02,0}$ when $f_\mathrm{target}=3.44$~GHz, and we focus on the emergence of the peaks in $\ket{01,1}$, which is the second- or third-most important transition of the adiabatic factor $D_{i,\dressedstate{02,0}}$ (Fig.~\ref{fig:D_second_order}(c)). We re-plot a zoomed-in and annotated version of the relevant simulations in Fig.~\ref{fig:1st2nd_order_leakage}. The previously described first-order effect coming from $\ket{11,0}$ leakage to $\ket{02,0}$ is illustrated by the broadest peak in Fig.~\ref{fig:1st2nd_order_leakage}(a) (blue line). We identify the emergence of non-uniform peaks in the $\ket{01,1}$ state (olive line in Fig.~\ref{fig:1st2nd_order_leakage}(a)). If we only consider first-order non-adiabatic effects, we expect leakage peaks at intervals of 0.94~ns that correspond to the $\dressedstate{11,0}\rightarrow\dressedstate{01,1}$ transition. This first-order transition does indeed line-up with a subset of the $\ket{01,1}$ peaks in the simulation (red vertical lines in Fig.~\ref{fig:1st2nd_order_leakage}(a)). If we now consider a potential second-order transition to $\ket{01,1}$ that originates from $\ket{02,0}$, we see that the additional set of expected peak intervals of 1.05~ns (blue vertical lines in Fig.~\ref{fig:1st2nd_order_leakage}(a)) match up with the remaining, previously unidentified peaks in the $\ket{01,1}$ state. This demonstrates that the leakage dynamics is better explained by considering both the first- and second-order bleeding effects. 

As further validation of the second-order bleeding effect, we look at the dips in the states from which the non-adiabatic transition originated. Dips at intervals of 0.94~ns from the first-order effect only show up in the $\ket{11,0}$ (red vertical lines match dips in the red trace of Fig.~\ref{fig:1st2nd_order_leakage}(b)). In comparison, dips at 1.05~ns intervals from the second-order effect do not appear in $\ket{11,0}$, and they only appear in $\ket{02,0}$ (blue vertical lines match dips in the blue trace of Fig.~\ref{fig:1st2nd_order_leakage}(a)). 

Here, we highlight that the leakage amplification method is intended to look at first-order bleeding effects, but under specific conditions, the second-order bleeding effect is also visible. More specifically, the peaks of the second-order bleeding will appear when the originating state is populated. This condition occurs when the time separation between the second-order peaks is within the width of the first-order peak. 
In our device, the predominant first-order bleeding channel is to $\ket{02,0}$, so the $\ket{02,0}$ exhibits peaks with the largest width in the leakage amplification simulation and, on our device, are the only peaks wide enough for any second-order peaks to be visible. Specifically we highlight the emergence of the $\ket{01,1}$ peaks in Fig~\ref{fig:1st2nd_order_leakage}(a), which happens only inside of the $\ket{02,0}$ peak. In summary, in terms of higher-order bleeding effects, the leakage amplification sequence helps amplify non-adiabatic transitions in the system beyond first-order bleeding, but it may not visualize all of the second-order effects that are present. In general, though, they demonstrate the complications that can arise in the regimes with large, competing $D$-factors. 

\subsection{Comparing symmetric versus asymmetric tunable coupler systems}
\label{sec:supplement_symVasym}
We run QuTiP simulations on a symmetric and asymmetric floating coupler system to compare the energy-level ordering and its implications. For the set of simulations presented in this section, we use qubit frequencies and anharmonicities $f_1=4.2\,$GHz, $f_2=4.3\,$GHz, $\eta_q/2\pi=-0.22\,$GHz; tunable coupler anharmonicity $\eta_c=-0.1\,$GHz; and $|g_{1c}|=|g_{2c}|=100\,$MHz. For the direct qubit-qubit coupling, we use slightly different values ($g_{12}=-6\,$MHz on the symmetric layout; $g_{12}=-7\,$MHz on the asymmetric layout) so that the residual $ZZ$ can be canceled out in the straddling regime for both sets of simulations. 

\begin{figure*}[tbp]
    \centering
    \includegraphics{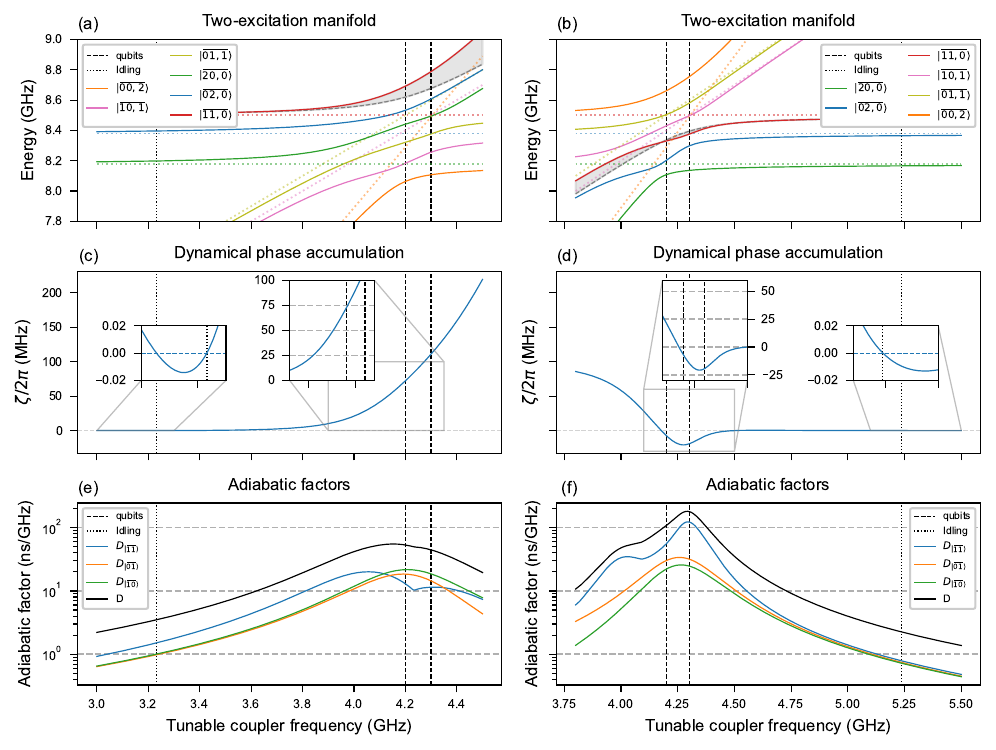}
    \caption{Simulated properties of the symmetric coupler (left column) and asymmetric coupler (right column) systems with comparable Hamiltonian parameters to ensure $ZZ=0$ is accessible. (a--b) Two-excitation manifold energy levels labeled with their closest bare eigenvector at idling. Black vertical dotted lines correspond to the coupler idling point, and black vertical dashed lines correspond to the qubit frequencies. The gray dashed line is $E_{10,0}+E_{01,0}-E_{00,0}$, with the gray area illustrating the dynamical phase accumulation. For the symmetric coupler in (a), $\dressedstate{11,0}$ has the highest frequency in the two-excitation manifold, while in (b), $\dressedstate{11,0}$ for the asymmetric system is bounded above and below by the other two-excitation energy levels. (c--d) Dynamical phase rate $\zeta$ represented by the gray area in the two-excitation manifold plots. Insets show a zoom-in of the regions of low $\zeta$ around $ZZ=0$ and the regions around the qubit frequencies, which would be a practical limitation on accessible gate speeds. When the couplers are pulsed from idling towards the qubit frequencies, the symmetric (asymmetric) coupler enables $\zeta/2\pi$ up to 75\,MHz (-20\,MHz). The $\zeta$-trajectory for the symmetric coupler keeps growing past the qubits while the $\zeta$ enacted by the asymmetric coupler eventually saturates. (e--f) Adiabatic factors for the computational states of the system summed over the contributions from coupled states $D_{11}$, $D_{01}$, and $D_{10}$ (colored lines) and the sum of these factors to give the total $D$ for the system (black line). As the couplers come close to the qubit frequencies, $D$-factor for the asymmetric coupler is almost an order of magnitude larger than for the symmetric coupler. 
    }
    \label{fig:comparison_sym_asym}
\end{figure*}

Because of the electrode configurations in a symmetric versus asymmetric coupler system \cite{Sete2021}, the two systems have a different idling point for zero coupling: symmetric (asymmetric) couplers idle at a lower (higher) frequency relative to the qubits and are pulsed higher (lower) in frequency to enact the gate. The two-excitation energy levels are presented in  Fig.~\ref{fig:comparison_sym_asym}(a--b), with a zoom-in of the $ZZ=0$ point included in the insets of Fig.~\ref{fig:comparison_sym_asym}(c--d). 

In terms of the coupler-mediated adiabatic gate, the key difference between the two types of coupler systems in the straddling regime is the order of the $\dressedstate{11,0}$ state relative to the other two-excitation energy levels at idling. In the case of the symmetric coupler, the energy configuration is favorable since $\dressedstate{11,0}$ is the highest frequency energy-level, which is not the case for the asymmetric coupler system (Fig.~\ref{fig:comparison_sym_asym}(a--b)). As a result, when the symmetric coupler is modulated upward in frequency from the idling point to enact the adiabatic CZ gate, $\dressedstate{11,0}$ is deflected only in the upward direction, resulting in an unbounded growth of the gray shaded area in Fig.~\ref{fig:comparison_sym_asym}(a) that translates to a unidirectional increase in the dynamical phase accumulation $\zeta_\mathrm{sym}$ shown in Fig.~\ref{fig:comparison_sym_asym}(c). 

In comparison, the downward frequency-trajectory of $\dressedstate{11,0}$ used to enact the gate in the asymmetric coupler system is more complicated since the other two-excitation energy levels are both above and below $\dressedstate{11,0}$ at idling and $\dressedstate{11,0}$ approaches and passes through various anticrossings during the trajectory, resulting in openings and closings of the gray area of the two-excitation manifold simulation  (Fig.~\ref{fig:comparison_sym_asym}(b)). In Fig.~\ref{fig:comparison_sym_asym}(d), this behavior translates to $\zeta_\mathrm{asym}$ that moves from negative to positive values. When the asymmetric coupler reaches the first qubit, the accumulated phase $\zeta_\mathrm{asym}/2\pi=-20\,$MHz is smaller in magnitude than $\zeta_\mathrm{sym}/2\pi=75\,$MHz that is attained on the symmetric system when the coupler reaches the first qubit (insets of Fig.~\ref{fig:comparison_sym_asym}(c--d)). Furthermore, as the coupler moves past the qubit frequencies, $\zeta$ eventually saturates to approximately 100$\,$MHz on the asymmetric system while $\zeta$ keeps increasing on the symmetric system. 

Besides gate speed, the energy-level ordering also has implications for the adiabatic factors. In Fig.~\ref{fig:comparison_sym_asym}(e--f), we plot the adiabatic factors for the different computational states of the qubits $D_k$ and the total $D$-factor that comes from summing the different $D_k$ contributions. The $D_{\dressedstate{11}}$-factor contributes the most for the early trajectory of the symmetric coupler and drops off close to the qubit frequencies, where the contributions from $D_{\dressedstate{01}}$ and $D_{\dressedstate{10}}$ begin to dominate, as shown in Fig.~\ref{fig:comparison_sym_asym}(e). In comparison for the asymmetric system (Fig.~\ref{fig:comparison_sym_asym}(f)), all three of the $D$-factors keep increasing as the coupler approaches the qubit frequencies. In this fast-gate regime, the $D$-factor for the asymmetric coupler system is almost an order of magnitude larger than for the symmetric coupler system and primarily comes from a substantial increase in the $D_{{\dressedstate{11}}}$ contribution. This is consistent with our earlier observations that the $\dressedstate{11}$ in the asymmetric system approaches or encounters more level crossings than in the symmetric system as the coupler is modulated away from idling to enact a gate. 

\section{Binomial distribution and maximum likelihood estimation of randomized benchmarking data}
Randomized benchmarking (RB) sequences are implemented by a series of random Clifford operations followed by a final Clifford to bring the system back to its initial state.
Each sequence of random Clifford is executed and measured $N_{\mathrm{shots}}$ times over $k$ different seeds, collecting a total of $kN_{\mathrm{shots}}$ shots.
Each shot is classified in the system computational states and the probability to bring the system back to the initial state is calculated. 

For a two-qubit RB sequence a binomial distribution is naturally generated by mapping the final result into the state $\ket{00}$ or not.
The statistical fluctuations of the data come from binomial sampling, not from Gaussian noise; its variance depends on the sample mean value at the sampled Clifford depth; and it shrinks asymmetrically near the boundaries of the distribution (see~\ref{sec:CI_binomial}).
Under these circumstances, maximum likelihood estimation is a better approach compared to least squares because it automatically incorporates the correct variance structure of the binomial distribution (see~\ref{sec:MLE}).

\subsection{Confidence intervals for binomial distributions}\label{sec:CI_binomial}
Wald and Wilson intervals are commonly used to estimate confidence intervals for binomially distributed data.
However, the choice of interval becomes important when dealing with small sample sizes or probabilities close to the physical boundaries (0 and 1), as is typical in high-fidelity RB experiments.

The Wald interval is the simplest confidence interval for binomial distributions.
It is derived from a normal approximation to the binomial distribution and produces a symmetric error bar around the estimated probability $\hat{p}$. The standard error is given by
$$
\sigma_{\mathrm{Wald}} = \sqrt{\frac{\hat{p}(1-\hat{p})}{N}}
$$
where $\hat{p}$ is the sample mean estimated from $N$ measurements.

The standard error is rescaled by the statistical $z$-score corresponding to the desired confidence level $C$.
Defining the significance level $\alpha = 1 - C$, the two-sided interval uses $z_{\alpha/2}$, which corresponds to a tail probability $\alpha/2$.
The 95\% confidence interval is
$$
CI_{\mathrm{Wald}}(95\%) = \hat{p} \pm z_{0.025} \, \sigma_{\mathrm{Wald}}
$$
with $z_{0.025} = 1.96$.

While straightforward to compute, the Wald interval performs poorly when $N$ is small or when $\hat{p}$ is close to 0 or 1.
In these regimes, the normal approximation becomes inaccurate and the resulting interval may extend beyond the physical range $[0,1]$.
This is particularly problematic for decay models or high-fidelity gates, where $\hat{p}$ is typically close to unity.

To ensure physically meaningful and statistically reliable confidence intervals, especially in the small-sample or near-boundary regime, we use the Wilson score interval,which leads to an asymmetric interval with improved coverage properties.

The 95\% Wilson confidence interval is given by
$$
CI_{\mathrm{Wilson}}(95\%) = \frac
{
\hat{p} + \frac{z_{0.025}^2}{2N} \pm z_{0.025} \sqrt{ \frac{\hat{p}(1-\hat{p})}{N} + \frac{z_{0.025}^2}{4N^2}}
}
{
1+\frac{z_{0.025}^2}{N}
}
$$
Other confidence levels are obtained by using the corresponding $z$-score. Importantly, the Wilson interval reduces to the Wald interval in the large-sample limit, but it maintains better coverage accuracy and remains within the physical range for finite $N$.

In this work, the error bars on averaged probabilities in interleaved RB measurements correspond to the 95\% confidence intervals estimated using the Wilson score method.

\subsection{Maximum Likelihood Estimation}\label{sec:MLE}
In a binomial distribution, the probability of measuring $k$ positive outcomes out of $N$ trials is
$$
L = \binom{N}{k}P^k(1-P)^{N-k}
$$
where $P$ is the theoretical probability distribution of the underlying model.
For two-qubit RB measurements, the positive outcomes are associated with the system being measured in $|00\rangle$, and the underlying model is $P(m) = Ap^m + B$, with $m$ denoting the Clifford depth and $p$ the depolarizing parameter.
The whole RB curve is sampled over $n$ discrete points $\{ m_i \}_{i\in[1,n]}$, with a total probability function
$$
L = \prod_{i=1}^n L_i = \prod_{i=1}^n \binom{N_i}{k_i}P_i^k(1-P_i)^{N_i-k_i}
$$
where the index $i$ represents values at the corresponding discrete sampling.

We define this as the likelihood function, but we will maximize the log-likelihood:
$$
l = \ln{(L)} = \sum_{i=1}^n\left[ k_i\ln{(P_i)} + (N_i-k_i)\ln{(1-P_i)} \right] + K
$$
where $P_i = P(m_i)$, and $K$ is a constant offset that can be removed in the optimization.

The parameters $\left( p, A, B \right)$ of the RB decay curve are defined in a constrained space $0 < \left( p,\; A,\; B \right) < 1$ and $A+B < 1$.
A direct optimization in this constrained space can lead to failure near boundaries (high fidelity gates) and make it difficult to compute and account for asymmetric error bars of raw data and derived fidelities.
For these reasons, it is preferred to run the maximum likelihood estimation in an unconstrained space by making use of link functions as unique maps from the physically bounded space to the unconstrained $\mathbb{R}$.
We will use the sigmoid function $\sigma(\theta)=\frac{1}{1+\exp(-\theta)} : \mathbb{R}\rightarrow (0,1)$ and its inverse $\text{logit}(\theta)=\ln\left( \frac{\theta}{1-\theta} \right): (0,1)\rightarrow\mathbb{R}$.
The constrained parameters $(p, A, B) \in (0,1)$ of the RB decay curve are mapped to the unconstrained $(\xi, \beta,\gamma)\in\mathbb{R}^3$ via the logit function:
\begin{align*}
    &\xi = \text{logit}(A) \\
    &\beta = \text{logit}(c) \\
    &\gamma = \text{logit}(p)
\end{align*}
with the variable $c=\frac{B}{1-A}$ introduced to automatically enforce the condition $A+B<1$.

With this new parametrization, the optimization via maximum likelihood estimation is run on an unconstrained space, making the optimization more robust and making it easier to compute the confidence intervals.
The optimized parameters and confidence intervals are finally converted back to the constrained space using the sigmoid function and Jacobian of the transformation.

This method accounts for statistical data fluctuations of the binomial sampling and their asymmetry near the boundaries, and it naturally constructs an asymmetric confidence interval of the optimized parameters in the physical constrained space.
By making use of Monte Carlo sampling, we can propagate asymmetric confidence intervals in $p$ to an asymmetric confidence interval in the two-qubit gate error from interleaved RB.
As we will see in the next section, a traditional error propagation method is sufficient for small and symmetric errors in $p$.

\subsection{Monte Carlo sampling for interleaved confidence intervals}
\label{sec:supplement_montecarlo}
The fidelity of the two-qubit CZ gate is calculated from the $p$ values of the reference and interleaved Clifford decay \cite{MagesanIRB2012}, using the equation:
\begin{equation}\label{eqn:error_from_RB}
r_{\mathrm{CZ}}=\frac{d-1}{d}\left(1-\frac{p_{\mathrm{iRB}}}{p_{\mathrm{RB}}} \right),
\end{equation}
where $d$ is the dimension of the two-qubit Hilbert space and $p_{\mathrm{RB}}$ and $p_{\mathrm{iRB}}$ are the depolarizing factors of the RB and interleaved RB sequences.
To estimate the error on $r_{\mathrm{CZ}}$ from the errors on $p$, a standard error-propagation formula is typically used. However, in the case of asymmetric error bars on $p$, Monte Carlo sampling is more appropriate for confidence interval estimation. 

We start by randomly sampling the normal distribution of the optimized parameters $\left( \xi, \beta, \gamma \right)$ in the unconstrained space for both the RB and interleaved data.
Mean values and covariance of the distributions are obtained from the MLE optimization.
The random samples of the unconstrained variables are mapped to the constrained parameters and the values of $p_{\mathrm{RB}}$ and $p_{\mathrm{iRB}}$ are estimated, from which we calculate the distribution of gate errors using eq.~\ref{eqn:error_from_RB} and remove non-physical values. The confidence interval $C$ on $r_\mathrm{CZ}$, which is not necessarily symmetric around the average value, is defined as the $\alpha/2$ and $1-\alpha/2$ quartiles of the distribution ($\alpha=1-C$ is defined in \ref{sec:CI_binomial}).
This method ensures the propagation of asymmetric error bars and avoids nonphysical confidence interval values.

\begin{figure}[bt]
    \centering
    \includegraphics[width=\columnwidth]{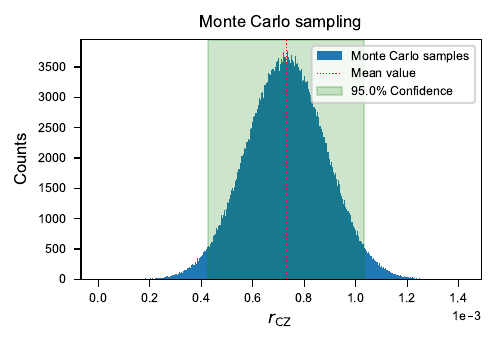}
    \caption{Monte Carlo sampling in the unconstrained space of the RB and interleaved RB curves, mapped to the constrained CZ gate error space. The 95\% confidence interval of the mean value is represented by the green area.}
    \label{fig:MonteCarlo}
\end{figure}
Fig~\ref{fig:MonteCarlo} is a Monte Carlo distribution over $5\times10^5$ randomly sampled $p$ values from the MLE analysis of the high fidelity gate in Fig.~\ref{fig:benchmarking}(a) of the main text.
The red dashed line at $7.2944\times10^{-4}$ is the distribution mean value, and the green box highlights the 95\% confidence interval $\left[4.261, 10.324 \right]\times 10^{-4}$.
The fidelity of the CZ gate has a 95\% confidence interval $\left[99.897, 99.957 \right]\%$, and an average value of 99.927\%.

We note that because of the low error bars in $p$ values, the Monte Carlo distribution is symmetric over the average value. As a result, a symmetric error bar on the gate fidelity $r_\mathrm{CZ}$ can be used.
The standard deviation $\sigma$ of the distribution is calculated by rescaling the confidence interval by the corresponding $z$-score (1.96 for a 95\% confidence interval).
This leads to the $r_{\mathrm{CZ}} = (0.073 \pm 0.015)\%$, or a gate fidelity of $(99.927 \pm 0.015)\%$. 

The standard deviation from the Monte Carlo distribution in this case is the same value as what comes from using the standard error propagation formula. So we highlight here, that because of the symmetric error bars in $p$, it is possible to apply the formula of error propagation, simplifying the process to estimate the error on the gate fidelity.
This is valid only if the errors on $p$ values are small compared to their value, ensuring symmetric confidence intervals.
All the two-qubit fidelities in this manuscript are quoted with a symmetric standard deviation $\sigma$ calculated with the error propagation formula from \ref{eqn:error_from_RB}, after verifying its agreement with the Monte Carlo analysis. 

\begin{figure}[ tb]
    \centering
    \includegraphics[width=\columnwidth]{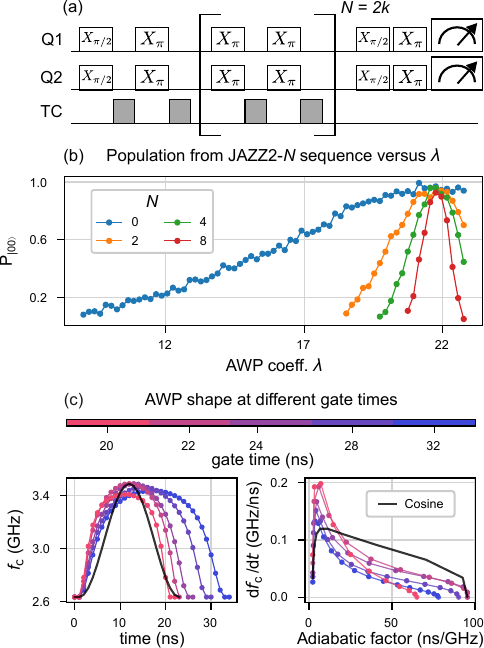}
    \caption{Tune-up procedure and results for the CZ gate enacted with an AWP shape. (a) JAZZ2-$N$ sequence from Ref.~\cite{GotoNakamura2024}, where $k$ is an integer value and $N$ corresponds to the number of coupler pulse repetitions. (b) An example of the JAZZ2-$N$ data used to find the $\lambda$ value that yields the maximum $P_{\ket{00}}$, which corresponds to the CZ gate operating point. (c) AWP shapes at different CZ gate times corresponding to the CZ gates benchmarked in Fig.~\ref{fig:fourierVawp}(b) of the main text. On the left, we plot the pulse in terms of $f_c$ versus time and on the right, we plot the speed of the pulse $\mathrm{d}f_c/\mathrm{d}t$ versus the total adiabatic factor. At higher values of $f_c$ and larger adiabatic factor values, the speed of the pulse slows down to maintain adiabaticity. Black lines represent a pure cosine pulse in terms of coupler frequency and are plotted as a reference for an unweighted pulse.}
    \label{fig:awpbringup}
\end{figure}
\section{Generation of the adiabatically-weighted pulse}
\label{sec:supplement_awptuneup}
To implement adiabatic CZ gates with fast gate times, the coupler frequency is moved close to the qubit frequencies to a point where dynamical $\zeta$ is large. However, in doing so, the system ends up in a regime where the hybridization between states is large and detuning between energy levels in the two-excitation manifold become small, as quantified by the adiabatic factor $D$ illustrated in Fig.~\ref{fig:sim_sym}(c). Since large $D$-factors indicate non-adiabatic transitions are likely, we aim to tailor the pulse shape so that the trajectory moves slowly when $D$ is large and moves quickly in zones with smaller $D$ (Eq.~\ref{eq:awp}). 

To tune-up the AWP, we perform the following steps: 
\begin{itemize}
    \item Estimate $D(\omega_c)$ from Eqs.~\ref{eq:dfactor} and \ref{eq:awp} in the main text using measured and design parameters. 
    \item Define a pulse in terms of $\omega_c$ by taking 
    \begin{equation}
        \omega_c(t)=G^{-1}\left(\frac{\lambda t_\mathrm{CZ}}{2\pi}\left[1-\cos\left(\frac{2\pi t}{t_\mathrm{CZ}}\right)\right]\right),
    \label{eq:supp_awpintegrate}
    \end{equation} where $G=\int_{\omega_0}^{\omega_c(t)}D(\omega')\mathrm{d}\omega'$, $\omega_0$ is the frequency of the coupler at idling, and $\lambda$ is the AWP weighting coefficient.
    \item Convert the pulse from $\omega_c(t)$ to pulse amplitude $V(t)$ using a fit to coupler spectroscopy data (Fig. \ref{fig:freqs_and_zz}(a)).
    \item Empirically find $\lambda$ using the pulse sequence from JAZZ2-$N$ (Fig.~\ref{fig:awpbringup}(a)). Instead of varying the flux amplitude, we sweep across values of $\lambda$ by regenerating Eq.~\ref{eq:supp_awpintegrate} for each value of $\lambda$. From data like the one shown in Fig.~\ref{fig:awpbringup}(b), we select a value of $\lambda$ that yields the maximum value of $P_{\ket{00}}$. 
\end{itemize}

In practice, we also utilize the Optuna optimizer \cite{optuna2019} at each gate time to fine-tune the final pulse shape, performing a search over three of the parameters that define $D(\omega_c)$ ($g_{12}$, $g_{1c}$, $g_{2c}$) and three of the parameters that define the coupler's frequency-to-amplitude conversion ($E_JE_C$, ratio of the JJs, and anharmonicity of the coupler). The benchmarking results from these Optuna runs were presented in Fig.~\ref{fig:fourierVawp}(b) of the main text. In Fig.~\ref{fig:awpbringup}(c), we also present the corresponding pulse shapes of the AWP at each gate time. As the gate times become shorter, the maximum $f_c$ in the pulse also becomes higher except for the shortest gate time of 20~ns, which has a lower fidelity (left plot of Fig.~\ref{fig:awpbringup}(c)). Furthermore, the speed of the gate is suppressed more as the total adiabatic factor becomes larger and vice versa (right plot of Fig.~\ref{fig:awpbringup}(c)). For reference, we plot an exemplary cosine shape in coupler frequency versus time with a maximum $f_c$ and pulse length corresponding to the 22~ns AWP shape. In this case, because there is no additional weighting factor, the speed of the pulse has a minimal decrease as the adiabatic factor (and likewise $f_c$) becomes large. 

\section{Leakage rate estimation}
\label{sec:supplement_leakage}
\begin{figure}[ tb]
    \centering
    \includegraphics[width=\columnwidth]{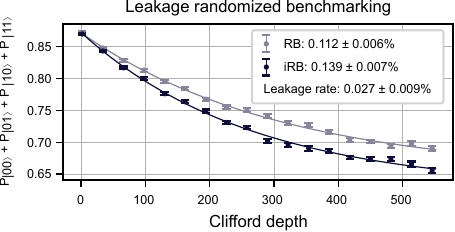}
    \caption{Leakage randomized benchmarking for the 24~ns AWP-shaped CZ gate measured for Fig.~\ref{fig:benchmarking}. From this measurement, we extract a leakage rate of 0.027\% per CZ. 
}
    \label{fig:leakage_rb}
\end{figure}

To estimate a leakage rate for the 24~ns gate reported in Fig.~\ref{fig:benchmarking} of the main text, we implement leakage randomized benchmarking following Ref.~\cite{Sung2021}. We perform standard RB and iRB measurements and calculate the population that remains in the computational subspace $P=P_{\ket{00}}+P_{\ket{01}}+P_{\ket{10}}+P_{\ket{11}}$, as shown in Fig.~\ref{fig:leakage_rb}. We fit the computational subspace population as a function of Clifford depth $m$ to $P(m)=Ap^m+B$, from which we can estimate an average leakage rate per Clifford $L_{\mathrm{(i)RB}} = (1 - B) (1 - p)$. The leakage rate per CZ is then given by $L_{CZ} = 1 - (1-L_{\mathrm{iRB}})/(1-L_{\mathrm{RB}})$. Using the same pulse parameters as those used for the gate reported in the main text, we estimate the leakage rate per CZ for the 24 ns AWP-shaped gate to be $0.027\pm 0.01\%$.

\section{Qubit coherence times}
\subsection{At the idling point}
\label{sec:supplement_cohIdling}
\begin{figure}[tb]
    \centering
    \includegraphics[width=\columnwidth]{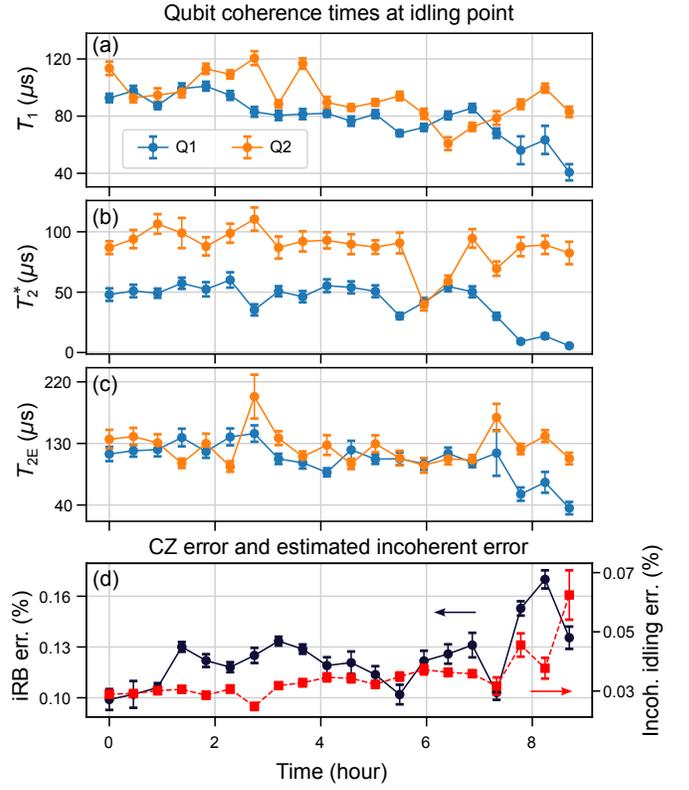}
    \caption{Time trace of qubit coherence times measured at the idling bias of the coupler and incoherent CZ error. Measurements come from the same 8.5-hour time frame as Fig.~\ref{fig:benchmarking}(b) in the main text. (a)-(c) Qubit coherence times $T_1$, $T_2^*$, and $T_\mathrm{2E}$ measured at the idling point of the coupler. (d) Comparing the CZ error from the main text of a 24-ns CZ gate+2-ns padding time (dark blue) with the estimated incoherent idling error $r_\mathrm{idling}$ (red). The estimated $r_\mathrm{idling}$ will be a lower bound on CZ error since the qubit coherence drops as the coupler moves towards qubits during the gate. The increase in CZ error during the time trace coincides with an increase in the idling incoherent error. }
    \label{fig:idlingCohs}
\end{figure}
We present the idling coherence times of Q1 and Q2 over the 8.5 hour time-trace taken during the benchmarking of a 24-ns adiabatic CZ gate (Fig.~\ref{fig:benchmarking}(a-b)). At idling during this period, the median coherence times for Q1 (Q2) are $T_1=81.4\,\mu$s ($91.2\,\mu$s), $T_2^*=49.7\,\mu$s ($89.5\,\mu$s), and $T_\mathrm{2E}=111.1\,\mu$s ($124.8\,\mu$s); the $T_2$ of Q2 starts to degrade towards the end of the run. We observe frequency beating in some of the $T_2^*$ measurements due to the relatively high charge-dispersion \cite{Koch2007} compared with the $T_2^*$ time (average measured charge-dispersion for Q1 and Q2 are approximately 17~kHz and 9~kHz and close to the calculated values).

Following Ref.~\cite{GotoNakamura2024}, we estimate the incoherent error $r_\mathrm{incoh}$ for an adiabatic CZ gate with interaction duration of $t_\mathrm{total}$ as 
\begin{equation}
\begin{aligned}
r_\mathrm{incoh}\simeq\sum_{q=\mathrm{Q1,Q2}} \left(\frac{t_\mathrm{total}}{5T_{1,q}}+\frac{2t_\mathrm{total}}{5T_{\mathrm{2E},q}}\right),
\label{eq:incoh_err}
\end{aligned}
\end{equation}
where we are neglecting the additional dephasing coming from qubit-qubit coupling during the gate that was considered in Ref.~\cite{GotoNakamura2024}. If we consider incoherent error from the idling coherence times of the qubit from Fig.~\ref{fig:idlingCohs}(a--c) and use $t_\mathrm{total}=28$ ns ($t_\mathrm{CZ}+2t\mathrm{pad}=24+2(2)$) as the total interaction time, we can estimate the total incoherent error at idling $r_\mathrm{idling}$ during the CZ stability run. This error rate provides a loose lower bound on coherence-limited fidelity because the effective coherence times will be lower when the TC frequency is increased to enact the gate (see Section~\ref{sec:supplement_cohFlux}). Though the measured CZ error is indeed higher than the estimated incoherent idling error, the rise in the $r_\mathrm{idling}$ approximation coincides with an increase in the measured CZ error at the end of the run, suggesting that the measured error is at least partially coherence-limited during portions of the time-trace.

\begin{figure}[tb]
    \centering
    \includegraphics[width=\columnwidth]{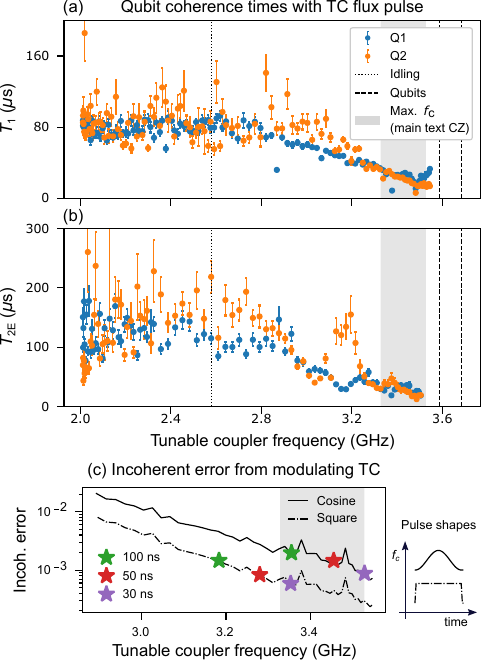}
    \caption{Impact of the coupler flux pulse on qubit coherence times. (a) $T_1$ and (b) $T_\mathrm{2E}$ of the qubits versus tunable coupler frequency. A square, dc pulse is applied to the coupler to move its frequency. Around the idling coupler frequency, $T_1$ and $T_\mathrm{2E}$ of the qubits start out around $80\,\mu$s and 100--150\,$\mu$s, respectively. As the coupler frequency moves closer to the qubits, qubit coherence times drop, with $T_1$ ($T_\mathrm{2E}$) dropping below 35 $\mu$s (40 $\mu$s) in the region where CZ gates from the main text are implemented (grayed out area). (c) Estimating incoherent error from qubit coherence times with coupler flux modulation for a cosine pulse (solid line) and square pulse (dashed-dotted line). Using a square pulse yields faster gate speeds at a given coupler frequency and therefore lower incoherent error. Frequencies for CZ gates with 30, 50, and 100~ns durations are marked for reference.}
    \label{fig:fluxCohs}
\end{figure}

\subsection{At the CZ operating point}
\label{sec:supplement_cohFlux}
To get a more accurate estimation of qubit coherence time during the gate, we look specifically at what happens to the qubit coherence times when the coupler frequency is pulsed toward the qubit frequencies. In Fig.~\ref{fig:fluxCohs}(a)-(b), we measure $T_1$ of the qubit with an additional square dc-pulse applied to the TC after the qubit's initial $\pi$-pulse and $T_{\mathrm{2E}}$ with a square dc-pulse applied to the TC between the $\pi/2$ pulses and refocusing $\pi$ pulse of the Hahn-echo sequence. 

As the coupler is pulsed towards the qubits, detuning between qubit and TC decreases and results in an increasing coupler component in the qubit dressed states. In the $T_\mathrm{2E}$ data, we observe that there is also an additional coherent interaction enacted near $f_c\approx3.2$ GHz, which we have not currently identified. In general, though, the qubit coherence times gradually decrease when $f_c\gtrsim2.8$, which signals that the qubits are becoming partially hybridized with a coupler with lower coherence times. As a result, for the CZ gate regime presented in the main text (shaded gray area), $T_1$ ($T_\mathrm{2E}$) of Q1 and Q2 ranges from approximately 35 $\mu$s down to 10 $\mu$s (40 $\mu$s down to 17 $\mu$s). These coherence times in the CZ gate regime are lower than the idling coherence times of approximately 80 $\mu$s (100--150 $\mu$s), which is likely why the Fourier cosine pulses scans showed improved fidelity for pulse envelopes that spend longer time near the idling point ($a_2<0$) (Fig.~\ref{fig:fourierVawp}(a)). 

With this data, we can estimate the incoherent error using the qubit coherence times measured when the coupler is modulated. We consider two pulse shapes defined in terms of $f_c$: a cosine pulse and a square pulse. We estimate the CZ gate time as a function of $f_{c,\mathrm{max}}$ by integrating the dynamical $\zeta(f_c)$ (Fig.~\ref{fig:freqs_and_zz}(b)) over the selected pulse shape from $f_\mathrm{idling}$ to $f_{c,\mathrm{max}}$ and selecting the gate time that yields $\phi_\mathrm{CPHASE}=\pi$. While the smoother cosine-like shape is necessary to reduce non-adiabatic transitions, it does result in a gate time that is slower than the gate time needed for a square pulse. (With the dynamical $\zeta$ rates of this device, the CZ gate time ends up being about 2.5--3.5 times slower with a cosine pulse than with a square pulse.) 

Using Eq.~\ref{eq:incoh_err}, we plot the estimated $r_\mathrm{incoh}$ for the two pulse shapes in Fig.~\ref{fig:fluxCohs}(c), marking the points that correspond to a 100~ns, 50~ns, and 30~ns CZ gate for reference. Though the qubit coherence times decrease with increasing coupler frequency, the gate time decreases quickly enough so that incoherent error from $T_1$ and $T_\mathrm{2E}$ of the qubits keeps going down. In general, the qubit coherence drop imposes a lower maximum CZ gate fidelity due to the higher incoherent error and highlights the importance of improving coupler coherence times at the 2Q gate operation point for high-fidelity adiabatic CZ gates.

\begin{figure}[ tb]
    \centering
    \includegraphics[width=\columnwidth]{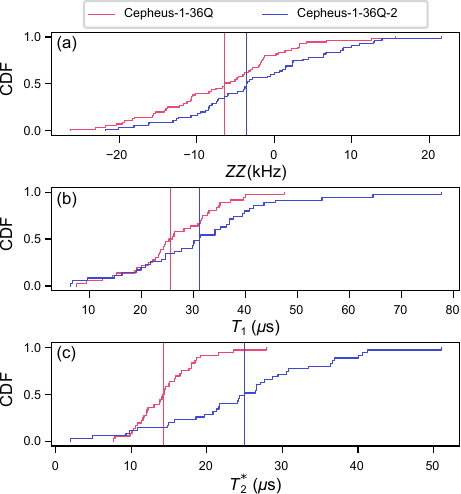}
    \caption{Cumulative distributions of the idling characteristics of the 36-qubit devices, Cepheus-1-36Q (magenta) and Cepheus-1-36Q-2 (teal). Median values for each device are marked by vertical lines. (a) Idling $ZZ$ for 57 (56) qubit-qubit edges and (b) $T_1$ and (c) $T_2^*$ for the 36 (35) qubits on Cepheus-1-36Q (Cepheus-1-36Q-2) devices. The median $ZZ$ magnitudes for both devices are less than 10kHz. Qubit coherence times are measured with couplers parked at their idling points near $ZZ=0$.}
    \label{fig:36Q_idling}
\end{figure}
\section{Adiabatic CZ gates on 36-qubit devices}
\label{sec:supplement_36q}
We look at the performance of fast adiabatic CZ gates on two 36-qubit devices---Cepheus-1-36Q$^\mathrm{TM}$ and Cepheus-1-36Q-2$^\mathrm{TM}$---each containing up to 60 qubit-qubit edges, and we report characterizations measured at a representative point in time. Both devices are composed of four 9-qubit chiplets connected through symmetric intermodular couplers (IMCs) \cite{Field2024}. Similar to the device studied in the main text, the qubit-qubit detuning within each 9-qubit chiplet is designed to be $\Delta_\mathrm{qq}\approx\alpha_q/2$. For IMC edges, the qubit-qubit detuning is designed to be $\Delta_\mathrm{qq}\approx 3\alpha_q/2$ as part of additional investigations on these 36-qubit devices. Despite the more restrictive frequency-allocation constraints associated with the increased qubit connectivity, the designed qubit-qubit detunings can be targeted even at the 36-qubit scale, as described in Ref.~\cite{XWang2026} together with additional frequency-allocation details for Cepheus-1-36Q. In Fig.~\ref{fig:36Q_idling}, we show cumulative distributions of the qubit coherence times and idling $ZZ$ values for Cepheus-1-36Q and Cepheus-1-36Q-2. The median $ZZ$ values are -6.4 kHz and -3.6 kHz, respectively. The median coherence times of Cepheus-1-36Q-2 are higher than those of Cepheus-1-36Q due to fridge variation. 

We first compare the performance of the adiabatic CZ gate with that of parametric-resonance gates \cite{Sete2021prgates} on Cepheus-1-36Q. For the adiabatic CZ gates, we use a raised cosine envelope for the coupler pulse with a fixed raised-cosine parameter:   
\begin{equation}
V(t)=
\begin{cases}
\frac{1}{2}\left[1-\cos\left(\frac{2\pi t}{\beta t_{\mathrm{CZ}}}\right)\right],
& 0 \le t < \tau_r, \\[6pt]
1,
& \tau_r \le t \le t_{\mathrm{CZ}}-\tau_r, \\[6pt]
\frac{1}{2}\left[1-\cos\left(\frac{2\pi(t_{\mathrm{CZ}}-t)}
{\beta t_{\mathrm{CZ}}}\right)\right],
& t_{\mathrm{CZ}}-\tau_r < t \le t_{\mathrm{CZ}} .
\end{cases}
\end{equation}
Here, the risetime $\tau_r=\beta t_{\mathrm{CZ}}/2$ and $0<\beta\le1$, and $\beta=0$ corresponds to a square pulse and $\beta=1$ corresponds to a cosine pulse so that the raised-cosine pulse spans a broader range of envelope shapes compared with the Fourier cosine pulse used on the main device. On Cepheus-1-36Q, we use a fixed value of $\beta=0.8$. The parametric-resonance gates are implemented by applying a modulated flux pulse to one qubit while simultaneously pulsing the coupler \cite{Sete2021prgates}.

\begin{figure}[ tb]
    \centering
    \includegraphics[width=\columnwidth]{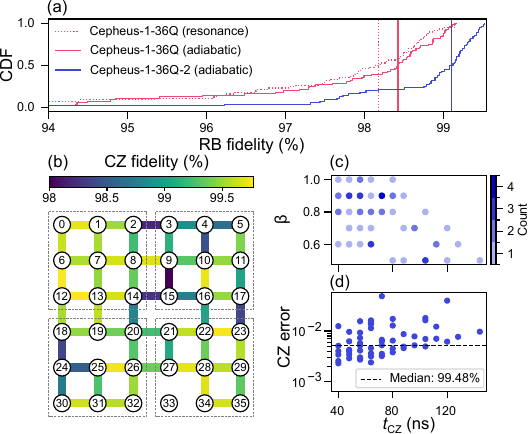}
    \caption{CZ gate performance on 36-qubit chiplet devices (a) Comparison of RB fidelities for parametric-resonance and adiabatic gates on Cepheus-1-36Q (magenta), where the adiabatic gates achieve a higher median fidelity of 98.42\% across 56 edges, compared with 98.17\% for the parametric-resonance gates across 57 edges. RB fidelities measured on a second device in a different fridge, Cepheus-1-36Q-2 (blue), have a median fidelity of 99.10\% across 56 edges. Median fidelities are marked by vertical lines. (b) Diagram of the 36-qubit chiplet device showing CZ fidelities measured on Cepheus-1-36Q-2 following an automated calibration procedure. Gray boxes indicate chiplet boundaries. (c) Raised-cosine pulse shapes used on Cepheus-1-36Q-2. Roll-off factors are roughly sampled from $\beta=0.5$ to $\beta=1$, where $\beta=1$ corresponds to a cosine pulse. The color intensity indicates the number of edges sharing the same $\beta$ and $t_{\mathrm{CZ}}$ values. (d) CZ error versus gate time for gates implemented on Cepheus-1-36Q-2. The median CZ fidelity on this device is 99.48\%.}
    \label{fig:36Q_gate}
\end{figure}
For each edge of Cepheus-1-36Q, we calibrate both parametric-resonance and adiabatic CZ gates using JAZZ2-$N$, benchmark their performance using standard randomized benchmarking, and report the average RB and CZ error per gate. Representative RB fidelities following a calibration run are shown in Fig.~\ref{fig:36Q_gate}(a). In particular, we highlight that on Cepheus-1-36Q, the adiabatic CZ gates achieve RB fidelities (median RB fidelity of 98.42\% across 56 edges) that are comparable to or better than the RB fidelities of the parametric-resonance gates (median RB fidelity of 98.17\% across 57 edges). 

As on the main-text device, modest pulse-shaping can further improve the performance of the adiabatic CZ gate. On Cepheus-1-36Q-2, we vary the raised-cosine parameter $\beta$ from 0.5 to 1 as part of an automated calibration procedure. The resulting RB fidelities are plotted in Fig.~\ref{fig:36Q_gate}(a) for comparison with Cepheus-1-36Q, and the distribution of the CZ fidelities across the device is shown in Fig.~\ref{fig:36Q_gate}(b). We note that Cepheus-1-36Q-2 has higher qubit coherence times than Cepheus-1-36Q and that sampling a range of $\beta$ values allows us to further improve the gate fidelities on Cepheus-1-36Q-2 relative to those obtained with a fixed pulse shape. The best-performing $\beta$ and gate times vary across the qubit-qubit edges, as shown in Fig.~\ref{fig:36Q_gate}(c). With this coarse pulse-parameter scan, the median CZ fidelity across 55 qubit-qubit edges reaches 99.48\%, with gate times down to 40~ns (Fig.~\ref{fig:36Q_gate}(d)). These results demonstrate that the adiabatic CZ gate can be calibrated efficiently across a 36-qubit device and that high fidelity can be achieved with only modest pulse-shaping.